\documentclass[final,5p,twocolumn,times]{elsarticle} 

\usepackage{caption}
\usepackage{graphicx}

\usepackage{subcaption}

\renewcommand{\baselinestretch}{1.0} 

\usepackage{latexsym}
\usepackage{amsmath}
\usepackage{amsfonts}
\usepackage{amssymb}
\usepackage{tikz}
\usetikzlibrary{trees}

\usepackage{comment}
\usepackage{enumitem}
\usepackage{booktabs}
\usepackage{wrapfig}
\usepackage{multirow}

\usepackage{siunitx}
\usepackage{longtable}

\usepackage{tikz,pgfplots}
\usepackage[customcolors]{hf-tikz}
\pgfplotsset{compat=1.13}
\usetikzlibrary{arrows,matrix,positioning}
\tikzset{mathterm/.style={draw=black,fill=white,rectangle,anchor=base}}
\tikzstyle{every picture}+=[remember picture]

\usepackage{lscape}
\usepackage{multimedia}
\usepackage{stmaryrd} 

\usepackage{multirow}

\usepackage{xspace}
\newcommand{\Rplus}{\protect\hspace{-.1em}\protect\raisebox{.35ex}{\smaller{\smaller\textbf{+}}}}
\newcommand{\Cpp}{\mbox{C\Rplus\Rplus}\xspace}

\journal{Computers \& Fluids}

\begin{document}

\graphicspath{{pics/}}

\begin{frontmatter}



\title{Wavelet-based parallel dynamic mesh adaptation for magnetohydrodynamics in the AMROC framework}


\author[inpe]{Margarete Oliveira  Domingues\corref{cor1}}
\ead{margarete.domingues@inpe.br}

\author[SOU]{Ralf Deiterding}
\ead{r.deiterding@soton.ac.uk}

\author[inpe]{Muller Moreira Lopes}
\ead{muller.lopes@inpe.br}

\author[ifsp]{Anna~Karina Fontes Gomes}
\ead{anna.gomes@ifsp.edu.br}

\author[inpe]{Odim~Mendes }
\ead{odim.mendes@inpe.br}

\author[mrs]{Kai~Schneider}
\ead{kai.schneider@univ-amu.fr}

\cortext[cor1]{Corresponding Author}
\address[inpe]{National Institute for Space Research (INPE), Av. Astronautas 1758 Jd. da Granja, S\~ao Jos\'e dos Campos, S. Paulo,  12.227-010, Brazil}

\address[SOU]{Aerodynamics and Flight Mechanics Research Group, University of Southampton, SO17 1BJ, United Kingdom}

\address[ifsp]{Instituto Federal de Educa\c{c}\~ao, Ci\^encia e Tecnologia de S\~ao Paulo,  C\^ampus Cubat\~ao, S. Paulo, Brazil}

\address[mrs]{Institut de Math\'ematiques de Marseille (I2M), Aix-Marseille Universit\'e, CNRS, Centrale Marseille, 39 rue Joliot-Curie, 13453 Marseille cedex 13, France}

\begin{abstract}

Computational magnetohydrodynamics (MHD) for space physics has become an essential area in understanding the multiscale dynamics of geophysical and astrophysical plasma processes, partially motivated by the lack of space data. 
Full MHD simulations are typically very demanding and may require substantial computational efforts. 
In particular, computational space-weather forecasting is an essential long-term goal in this area, motivated for instance by the needs of modern satellite communication technology.
We present a new feature of a recently developed compressible two- and three-dimensional MHD solver, which has been successfully implemented into the parallel AMROC (Adaptive Mesh Refinement in Object-oriented C++) framework  with improvements concerning the mesh adaptation criteria based on wavelet techniques.
The developments are related to computational efficiency while controlling the precision using dynamically adapted meshes in space-time in a fully parallel context.

\end{abstract}


\end{frontmatter}
\section{Introduction}

Space weather forecasting is concerned with predicting disturbances in the Earth upper atmosphere and the magnetic field that can have dramatic consequences for modern technology, especially satellites and communication electronics. 
A combination of theoretical studies, observations, numerical simulations including data assimilation, and forecasting is the key for achieving success in this strategic area \cite{Schrijver20152745}. 
For numerical predictions of space plasma physics, the considerable range of dynamically active space-time scales is a major obstacle. 
Accordingly, fast, robust, and efficient numerical models that merge physics-based, accurate simulation with timely observations are of fundamental importance. 
A particularly successful computational approach is the magnetohydrodynamic model \cite{Toth2012870}. 
Roughly speaking, the magnetohydrodynamic model consists of a system of eight nonlinear partial differential equations describing the dynamics of a compressible, inviscid, 
and perfectly electrically conducting fluid interacting with a magnetic field, combining thus the Euler equations of hydrodynamics with the Maxwell equations of electrodynamics. 
The latter yields an evolution equation for the magnetic field, the so-called induction equation, and an incompressibility constraint of the magnetic field using Gauss's law \cite{Bittencourt:2004}.
Numerical approaches to solve these systems are computationally costly and a full mesh approach to compute them is prohibitive in most cases of interest in space physics \cite{Toth2012870}. 
Therefore, adaptive techniques have been combined with on-the-fly mesh refinement \cite{Toth2012870}.  
More recently, multiresolution techniques have also been used \cite{Gomes2015199,DominguesetalESAIM:2013}. 
In such approaches, the mesh is refined locally just in the regions where structures or discontinuities are or can be present in short integration time \cite{Hejazialhosseini20108364}.

Adaptive techniques reduce the computing time significantly while preserving the high accuracy of the numerical solutions.
In a previous article \cite{DeiterdingDominguesGomesRousselSchneiderESIAM:2009} 
we presented a comparison of these mesh adaptation techniques, which we have recently extended  in \cite{Deiterdingetal2016SIAM}.
In these works, we used the generic open-source framework for \textcolor{black}{patch}-structured adaptive mesh refinement AMROC \cite{Deiterding-PhDThesis, DeiterdingESAIM:2011}. 
A core finding of these publications is that the multiresolution approach is mathematically more rigorous and leads to a more faithful mesh adaptation; yet, the \textcolor{black}{patch}-based adaptation approach -- thanks to advantageous data structures -- reduces the overall computation time drastically. 
Hence, in this work we present the multiresolution approach in our MHD solver \cite{MoreiraLopesetal2018CAF} within the AMROC framework.

The organisation of the remainder of the manuscript is as follows.
In Section~\ref{sec:environment}, we contextualise the space environment of the Earth and introduce the primary phenomena of interest for simulation test cases. 
In Section~\ref{secAMR} we describe the governing equations, their discretisation using finite volumes, and outline the main ideas of the \textcolor{black}{patch}-based adaptive mesh refinement (AMR) approach briefly as implemented in AMROC. 
In Section~\ref{secExperiments} we develop numerical experiments and discuss the accuracy and efficiency of our implementation. 
Finally, in Section~\ref{secConclusions} we draw some conclusions.

\section{Earth space environment} \label{sec:environment}

The Sun is the source of several phenomena that affect the sidereal bodies, even the human artefacts, such as interplanetary probes, existing in the heliosphere~\cite{Kallenrode:2004}.
The three main solar agents are electromagnetic radiation, high energetic corpuscular radiation, and magnetised plasma structures evolving in the solar wind.

As the Earth has an atmosphere and an intrinsic magnetic field, a particular situation occurs; the solar radiation ionises the upper atmosphere of the Earth, located above an altitude of $70~\mathrm{km}$.
Also, a magnetic field of inner origin imprecates the whole atmosphere.
This conjunction of ionisation and magnetic field creates a kind of shield to the solar plasma displacement, i.e., the fully ionised and magnetised solar wind plasma cannot mix with the terrestrial plasma, which establishes a geomagnetic field domain surrounding the planet, named  magnetosphere.
As a consequence, the expanding solar plasma deviates from its original direction involving the Earth's domain~\cite{KivelsonRussell:1996}.

Incident upon this obstacle, the solar plasma wind, which moves with supersonic speed, creates a bow shock involving the terrestrial domain.
Immediately after the shock the flux presents a subsonic speed, thermalised particles, and an intensified interplanetary magnetic field, characterising a region designated as magnetosheath.
The plasma of this region compresses the region dominated by the geomagnetic field in a process that defines an interface between the two physical media, the magnetopause.
As a manifestation of a tangential discontinuity, this region is a surface of total pressure equilibrium between the solar wind-magnetosheath plasma and the geomagnetic field confined in the magnetosphere.
Inside the magnetosphere, several processes establish distinct regions of plasma, energetic particle distributions, and a sophisticated building of electrical current systems.
All the features of this real scenario are consequences of the electrodynamical interaction between the incident solar wind and the Earth atmosphere~\cite{Cravens:2004}.

An enormous amount of studies provided by experimental, observational research, such as the one obtained by satellite data-set analyses, and theoretical approaches, such as the propositions of phenomena from the magnetohydrodynamic formalism, has contributed to an in-depth interpretation of the space environment~\cite{RussellEtAl:2016, BaumjohannTreumann:1999}.
The efforts of numerical simulations can significantly help the scientific concepts under development and provide more realistic process descriptions.
The comprehensive view and understanding of important space plasma processes and their geoeffective events 
depend on much more realistic performances of the currently available magnetohydrodynamic models.
However, to be utilised in emerging space weather programs, competitive-in-time codes are increasingly demanded nowadays.

In this work, we aim at developing an \textcolor{black}{up-to-date} numerical  tool for modelling magnetohydrodynamics. 
We address two fundamental cases, prominent in the investigation of applied space sciences.
The first case is the Orszag--Tang vortex, originally introduced  in \cite{OrszagTang:1979} for incompressible MHD, a canonical model problem for testing the transition to supersonic two-dimensional MHD turbulence.
The second is the magnetic shock cloud~\cite{RussellEtAl:2016,BaloghTreumann:2013}, a common occurrence evolving superimposed with the solar wind through the interplanetary medium.

\subsection{GLM-MHD model}
In  numerical simulations of the ideal magnetohydro\-dynamics equations,
including the divergence-free constraint of the magnetic field, typical methodologies consider an additional correction term to facilitate the enforcement of this physical property. There are different kinds of well-known methods in order to minimise this effect, as described in \cite{Hopkins:2016} and references therein.  
In the context of this study, having in mind the application of the multiresolution method  using explicit time integration, we adopt the approach  proposed in \cite{Dedneretal:2002}, with the non-dimensional adjustment added by Mignone and Tzeferacos in \cite{mignone2010second}.
Namely,  the divergence-free constraint is treated  by the introduction of a new variable  $\psi$ and a corresponding balance equation is added to the ideal MHD equations. 
This process leads to the well-known Generalised Lagrangian Multiplier (GLM) hyperbolic conservation system 
\begin{equation}\label{eq:glmmhd}
	  \begin{aligned}
	     \displaystyle\frac{\partial\rho}{\partial t}+\nabla\cdot \left(\rho {\bf u}\right)&=&0,\\
		\displaystyle\frac{\partial \left(\rho {\bf u}\right)}{\partial t} + \nabla\cdot\left[\rho{\bf uu^T} + {\bf I}\left( p + \frac{{\bf B.B}}{2} \right)-{\bf BB^T}\right]&=& \mathbf{0},\\
		\displaystyle\frac{\partial{\bf B}}{\partial t} + \nabla\cdot\left[{\bf uB^T-Bu^T}+ \psi {\bf I}\right]&=& \mathbf{0}, \\
		\displaystyle\frac{\partial E}{\partial t} +\nabla\cdot\left[\left(E + p + \frac{{\bf B.B}}{2}\right){\bf u} - {\bf B} \left({\bf u.B} \right)\right] &=& 0,\\
	      \frac{\partial\psi}{\partial t}+c_h^2\nabla\cdot{\bf B}&=&-\frac{c_h^2}{c_p^2}\psi,
  \end{aligned}
\end{equation}
where $\rho$ represents density, $p$ is the pressure, ${\bf u}$ is the fluid velocity vector, ${\bf B}$ is the magnetic field vector, ${\bf I}$ is the identity tensor of order $2$, 
and the superscript symbol $T$ indicates  the transposed matrix. 
The parameter $c_h$ is defined as
$ c_h = \max\left[\nu\frac{\Delta h}{\Delta t},\,\max\left( |u_i|\pm c_f \right)\right],$
where $\Delta h$ is the minimal value of the mesh sizes in each direction, $\nu$ the Courant number, $u_i$ is the velocity of the $i$-th component, and $c_f$ is the fast magneto-acoustic wave of the MHD model. 
The $c_p$ value is defined in terms of the parameter $\alpha_p=\Delta h \, \dfrac{c_h}{c_p^2}$, where $\alpha_p\in[0,1]$, as described in \cite{mignone2010second}.
The total energy density $E$ is given by the constitutive law
	\begin{equation}
		\displaystyle E=\frac{p}{\gamma-1}+\rho\frac{{\bf u\cdot u}}{2}+\frac{{\bf B\cdot B}}{2}\,,
\label{eq:pressure}
	\end{equation}
in which $\gamma$ is the adiabatic constant with  $\gamma>1$.
Moreover, the above MHD system is completed by suitable initial and boundary conditions as presented in the numerical experiments section. These equations have been rewritten in non-dimensional form such that the magnetic permeability yields the identity, i.e., $\mu=1$. 
We also consider the divergence control parameter $\alpha_p=0.4$, and $\psi\equiv0$ in the initial conditions.

In near Earth space, the governing equations of MHD modelling can in particular develop discontinuities, i.e., shocks and contact waves. 
Therefore, we use finite volume shock-capturing methods that are constructed to properly handle this behaviour in a robust and oscillation-free way, as discussed in detail in \cite{Leveque:2002}. 
Moreover, we are interested in studying the development of MHD instabilities that are very local and can present
complex local multiscale  behaviour.
For this reason, it can be undoubtedly beneficial to have an economical, accurate, and efficient mesh representation of these features. 

\section{Numerical discretisation} 
\label{secAMR}

As reference discretisation for these equations in the conservation form, the numerical solution is represented by the quantity vector $\mathbf{Q}$ of the approximated cell averages on a uniform mesh of the computation domain. 
For space discretisation, a finite volume method is chosen, which results in a system of ordinary differential equations with a vector of numerical flux function differences with respect to each cell. 
In all numerical schemes throughout this paper enhanced numerical flux functions with comparable second-order-accurate reconstruction and
flux limiting are used. 
For time integration, we adopt an explicit second order Runge--Kutta scheme.

The adaptive mesh refinement method (AMR) \cite{Berger-Oliger-84,Berger-Collela-88,Bell-Berger-Saltzman-94} follows a patch-oriented refinement approach, where
non-overlapping rectangular submeshes 
$G_{\ell,m}$ define the domain  $G_\ell := \bigcup_{m=1}^{M_\ell} G_{\ell,m}$ of an entire level $\ell=0,\dots, L$.
As the construction of refinement proceeds recursively, a hierarchy of submeshes 
successively contained within the next coarser level domain is created. 
The recursive nature of the algorithm only allows the addition of one new level in each refinement operation.  
The patch-based approach does not require special
coarsening operations; submeshes are simply removed from the hierarchy. 
The coarsest possible resolution is thereby restricted to the level $0$ mesh. 
Typically, it is assumed that all mesh widths on level $\ell$ are $r_\ell$-times finer 
than on the level $\ell-1$, \textit{i.e.},  $\Delta t_\ell=\Delta t_{\ell-1}/r_\ell$ and $\Delta x_{n,\ell}=\Delta x_{n,\ell-1}/r_\ell$, with 
$r_\ell\in\mathbb{N}, r_\ell\ge 2$ for $\ell>0$ and $r_0=1$. This ensures that a time-explicit finite volume scheme remains stable under a CFL-type condition on all levels of the hierarchy. \textcolor{black}{In our MR implementations we always use $r_\ell=2$ here.}

The numerical update is applied on the level $\ell$ by calling a single-mesh routine implementing the finite volume scheme in a loop over all the submeshes $G_{\ell,m}$. The regularity of the input data allows a straightforward implementation of the scheme and furthermore permits optimisation to take advantage of high-level caches, pipelining, etc. 
New refinement meshes are initialised by interpolating the vector of conservative quantities ${\bf Q}$ from the next coarser level. 
However, data in an already refined cell are copied directly from the previous refinement patches. {\it Ghost}  cells around each patch 
are used to decouple the submeshes computationally. 
Ghost cells outside of the root domain $G_0$ are 
used to implement physical boundary conditions. Ghost cells in $G_\ell$ have a unique interior cell analogue and are set by copying the data value from the patch where the interior cell is contained (synchronisation). 
For $\ell>0$, internal boundaries can also be used. 
If recursive time step refinement
is employed, ghost cells at the internal refinement boundaries on the level $\ell$ are set by time-space interpolation from the two previously calculated time steps of level $\ell-1$. Otherwise, spatial interpolation from the level $\ell-1$ is sufficient.

One feature of the AMR algorithm is that refinement patches overlay coarser mesh data structures instead of being embedded, again preventing data fragmentation. Values of cells covered by finer submeshes are subsequently overwritten by averaged fine mesh values, which, in general, would lead to a loss of conservation on the coarser mesh. A remedy to this problem is to replace the coarse mesh numerical fluxes at refinement boundaries with the sum of fine mesh fluxes along the corresponding coarse cell boundary. 

\subsection{MR refinement indicator}\label{sec:criteria}

The principle of MR methods is the transformation of the cell averages given by the finite volume discretisation into a multiscale representation.
A detailed review can be found in \cite{DominguesetalESAIM:2011,mueller:2001} and references therein. 
\textcolor{black}{
We consider a discrete solution of the discretisation as initial cell average data $\bar{\mathbf{Q}}^{\ell+1}$ at level $\ell +1$.
Then for instance, in one decomposition level a two-level MR transformation can be written as follows,
\[
 \bar{\mathbf{Q}}^{\ell+1} \; {\overset{projection}
{\underset{prediction}{\rightleftharpoons}}} \; \bar{\mathbf{Q}}^{\ell+1}_{\text{MR}} \; = \;
\{\bar{\mathbf{Q}}^{\ell}\}\ \cup \ \{d^{\ell}\},
\]
where $d^\ell$ contains the information between the two consecutive levels $\ell$  and $\ell+1$;  and $\bar{\mathbf{Q}}^\ell$ stores a smooth version of the original 
numerical solution $\bar{\mathbf{Q}}^{\ell +1}$. 
}
These ideas are a natural extension of the work \cite{Harten:1995,Harten:1996}. 
The numerical solution at the finest resolution level is transformed  into a set of coarser scale approximations plus a series of prediction errors corresponding to wavelet coefficients. 
These coefficients describe the difference between subsequent resolutions. 
The main \textcolor{black}{principle} is then to use the decay of the wavelet coefficients to estimate the local regularity of the solution \cite{CohenKaberMUllerPostel:2003,Cohen:2000}.
In regions where the solution is smooth these coefficients are small, while they have large magnitude in regions of steep gradients or discontinuities.

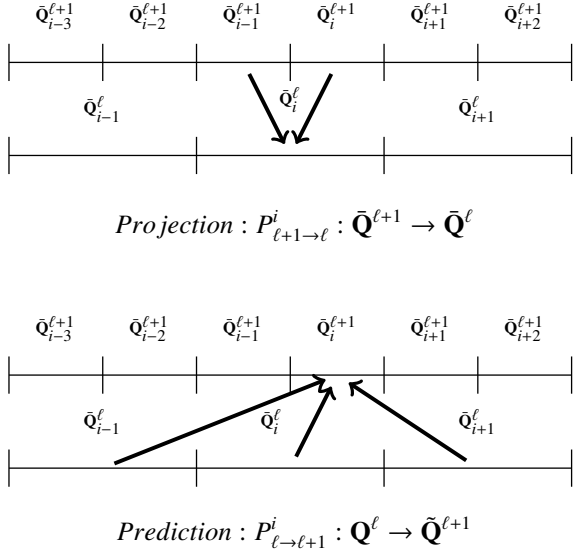
\begin{figure}[htb]
\hfsetfillcolor{gray!5}
\hfsetbordercolor{gray}
\begin{center}
\newcommand*{\xMin}{0}%
\newcommand*{\xMax}{6}%
\newcommand*{\yMin}{-1}%
\newcommand*{\yMax}{2}%

\resizebox{0.85\linewidth}{!}{%
\begin{tikzpicture}
\path (0.5,1.5) node (idn0) [ ] {{\tiny $\bar{\mathbf{Q}}^{\ell+1}_{i-3}$}};
\path (1.5,1.5) node (idn1) [ ] {{\tiny $\bar{\mathbf{Q}}^{\ell+1}_{i-2}$}};
\path (2.5,1.5) node (idn2) [ ] {{\tiny $\bar{\mathbf{Q}}^{\ell+1}_{i-1}$}};
\path (3.5,1.5) node (idn3) [ ] {{\tiny $\bar{\mathbf{Q}}^{\ell+1}_i$}};
\path (4.5,1.5) node (idn4) [ ] {{\tiny $\bar{\mathbf{Q}}^{\ell+1}_{i+1}$}};
\path (5.5,1.5) node (idn5) [ ] {{\tiny $\bar{\mathbf{Q}}^{\ell+1}_{i+2}$}};

\draw (0,1) -- (6,1);
\foreach \x in {0,...,6}
	\draw (\x,0.8)--(\x,1.2);

\path (1,0.5) node (idn6) [ ] {{\tiny $\bar{\mathbf{Q}}^{\ell}_{i-1}$}};
\path (3,0.6) node (idn7) [ ] {{\tiny $\bar{\mathbf{Q}}^{\ell}_{i}$}};
\path (5,0.5) node (idn8) [ ] {{\tiny $\bar{\mathbf{Q}}^{\ell}_{i+1}$}};

\draw (0,0) -- (6,0);
\foreach \x in {0,...,3}
	\draw (2*\x,-0.2)--(2*\x,0.2);

\path (3.5,1) node (l3k) [ ] {};
\path (2.5,1) node (l3kM1) [ ] {};
\path (3,0) node (l2k) [ ] {};

\draw[->,very thick] (l3k) -- (l2k);
\draw[->,very thick] (l3kM1) -- (l2k);
\end{tikzpicture}
}

\end{center}
\[
Projection: 
P^i_{\ell+1 \rightarrow \ell}: {\bar{\mathbf Q}}^{\textcolor{black}{\ell+1}} \rightarrow {\bar{\mathbf Q}}^{\textcolor{black}{\ell}} 
\]
\hfsetfillcolor{gray!5}
\hfsetbordercolor{gray}
\begin{center}
\newcommand*{\xMin}{0}%
\newcommand*{\xMax}{6}%
\newcommand*{\yMin}{-1}%
\newcommand*{\yMax}{2}%

\resizebox{0.85\linewidth}{!}{%
\begin{tikzpicture}

\path (0.5,1.5) node (idn0) [ ] {{\tiny $\bar{\mathbf{Q}}^{\ell+1}_{i-3}$}};
\path (1.5,1.5) node (idn1) [ ] {{\tiny $\bar{\mathbf{Q}}^{\ell+1}_{i-2}$}};
\path (2.5,1.5) node (idn2) [ ] {{\tiny $\bar{\mathbf{Q}}^{\ell+1}_{i-1}$}};
\path (3.5,1.5) node (idn3) [ ] {{\tiny $\bar{\mathbf{Q}}^{\ell+1}_i$}};
\path (4.5,1.5) node (idn4) [ ] {{\tiny $\bar{\mathbf{Q}}^{\ell+1}_{i+1}$}};
\path (5.5,1.5) node (idn5) [ ] {{\tiny $\bar{\mathbf{Q}}^{\ell+1}_{i+2}$}};

\draw (0,1) -- (6,1);
\foreach \x in {0,...,6}
	\draw (\x,0.8)--(\x,1.2);

\path (1,0.5) node (idn6) [ ] {{\tiny $\bar{\mathbf{Q}}^{\ell}_{i-1}$}};
\path (2.8,0.5) node (idn7) [ ] {{\tiny $\bar{\mathbf{Q}}^{\ell}_{i}$}};
\path (5,0.5) node (idn8) [ ] {{\tiny $\bar{\mathbf{Q}}^{\ell}_{i+1}$}};

\path (3.5,1) node (l3k) [ ] {};
\path (1,0) node (l2kM1) [ ] {};
\path (3,0) node (l2k) [ ] {};
\path (5,0) node (l2k1) [ ] {};

\draw (0,0) -- (6,0);
\foreach \x in {0,...,3}
	\draw (2*\x,-0.2)--(2*\x,0.2);
\draw[->,very thick] (l2kM1) -- (l3k);
\draw[->,very thick] (l2k) -- (l3k);
\draw[->,very thick] (l2k1) -- (l3k);
\end{tikzpicture}
}

\end{center}
\[
Prediction: P^i_{\ell \rightarrow \ell+1}: {\bf Q}^{\textcolor{black}{\ell}} \rightarrow \tilde{{\bf Q}}^{\textcolor{black}{\ell+1}}
\]

\caption{
\label{fig:schemePP}
\textcolor{black}{Scheme of the projection (restriction) and prediction (prolongation) operators for the quantity vector $\bar {\mathbf Q}$.}}
\end{figure}

In order to perform the MR method, some operations for projection and prediction are required. 
For the MR scheme with finite volumes, where the cell values are local averages, a coarser cell $\Omega^\ell_i$ has its value estimated using the smaller scale values and a unique projection operator $P_{\ell+1 \rightarrow \ell} \, : \, \bar{\mathbf{Q}}^{\ell+1} \, \mapsto \, \bar{\mathbf{Q}}^{\ell}$. 
In this scheme, the projection operator to obtain the solution on a coarser cell is given by the average value of its children. 
For the one-dimensional case \textcolor{black}{(cf. Fig.~\ref{fig:schemePP}, top)}, the projection is performed by
\begin{equation}
    \bar{\mathbf{Q}}^\ell_i = 
P_{\ell+1 \rightarrow \ell} \left(\bar{\mathbf{Q}}^{\ell+1}_{2i}, \bar{\mathbf{Q}}^{\ell+1}_{2i+1}\right) = 
\frac{1}{2}\left( 
\bar{\mathbf{Q}}^{\ell+1}_{2i} + \bar{\mathbf{Q}}^{\ell+1}_{2i+1}\right),
\end{equation}
where $\bar{\mathbf{Q}}^{\ell}_{i}$ is the average value of the cell $\Omega^\ell_i$. 

The prediction operators are used to perform the opposite path of the projection operators and allow to obtain the values of the finer cells using the values of the coarser ones \textcolor{black}{(cf. Fig.~\ref{fig:schemePP}, bottom)}. 
For each child cell \textcolor{black}{$i$} to be predicted, there is a different prediction operator represented by $P^i_{\ell \rightarrow \ell+1} \, : \, \bar{\mathbf{Q}}^{\ell} \, \mapsto \, \bar{\mathbf{Q}}^{\ell+1}$ for the one-dimensional case.
These operators yield a non-unique approximation of $\bar{\mathbf{Q}}^{\ell+1}_i$ by interpolation. 

We use polynomial interpolation of second degree on the cell-averages as proposed by Harten~\cite{Harten:1995}, which yields third-order accuracy. 
For the one-dimensional case, it follows that
\begin{equation}
\tilde{\mathbf{Q}}^{\ell+1}_{2i}  = P^0_{\ell \rightarrow \ell+1} \left(\bar{\mathbf{Q}}^{\ell}_{i-1}, \bar{\mathbf{Q}}^{\ell}_{i}, \bar{\mathbf{Q}}^{\ell}_{i+1}\right) = \bar{\mathbf{Q}}^\ell_{i}- \frac{1}{8} (\bar{\mathbf{Q}}^\ell_{i+1} - \bar{\mathbf{Q}}^\ell_{i-1})
\end{equation}
\begin{equation}
\tilde{\mathbf{Q}}^{\ell+1}_{2i+1}  = P^1_{\ell \rightarrow \ell+1} \left(\bar{\mathbf{Q}}^{\ell}_{i-1}, \bar{\mathbf{Q}}^{\ell}_{i}, \bar{\mathbf{Q}}^{\ell}_{i+1}\right) = \bar{\mathbf{Q}}^\ell_{i}+ \frac{1}{8} (\bar{\mathbf{Q}}^\ell_{i+1} - \bar{\mathbf{Q}}^\ell_{i-1}),
\end{equation}
where $\tilde{\mathbf{Q}}^\ell_{i}$ is an approximation of the value $\bar{\mathbf{Q}}^{\ell}_{i}$. With this choice, the operator satisfies the locality and the consistency properties, namely:
 the interpolation into a child cell is computed from the cell-averages of its parent and its nearest uncle cells in each direction; and prediction and projection operator are consistent,  \textit{i.e.} $P_{\ell+1 \rightarrow \ell} \circ P_{\ell \rightarrow \ell+1} = \mbox{Identity}$.

The prediction operator is used to obtain the wavelet coefficients $d^\ell_i$ of the finer cells. 
The wavelet coefficients are then given by the difference between the values on the finer level and the predicted values as
\begin{equation}
    d^\ell_i = \bar{\mathbf{Q}}^{\ell}_{i} - \tilde{\mathbf{Q}}^{\ell}_{i}.
\end{equation}
The values $d^\ell_i$ are also used for reconstructing the finest levels without errors due to their property of being the interpolation error. 
Their norm yields the local approximation error. 
Moreover, the information of the cell-average value of the two children is equivalent to the knowledge of the cell-average value of the parent and one independent detail. 

The same idea can be extended to higher-dimensional cases. 
For instance, for two dimensions the information of the cell-averages of four children is equivalent to the knowledge of three wavelet coefficients in the different directions and the nodal value on the coarser mesh.

In the process to obtain the adaptive meshes, we flag all cells in which the associate wavelet coefficients $d^\ell$ are larger than a threshold.
We can select $d^\ell$ for only a scalar value from the MHD quantities or adopt other combinations between the variables.
For the examples presented below we consider only a scalar value for density or pressure.

\subsubsection*{Choice of the threshold}

There are different possible choices for the threshold, which enable the identification of the retained wavelet coefficients having magnitude above the threshold. 
In practice, its value should be chosen such that the perturbation related to the thresholding and the discretisation errors is of the same order. 
Moreover, it is possible, for instance, to use a constant threshold value $\epsilon$ for all levels. 
However, in the context of finite volumes, and based on our own experiences, \cite{DominguesGomesRousselSchneider:APNUM2009,DominguesetalESAIM:2013}, we usually follow Harten's thresholding strategy, i.e,
\begin{equation}
\epsilon^\ell=\frac{\epsilon}{|\Omega|} 2^{d\,\left(\ell-\mathbb{L}\right)}, \;\;1\leq \ell \leq \mathbb{L},
\end{equation}
in order to control the $L_1$-norm.
In this case,  $\mathbb{L}$ is the finest scale level added to the base mesh level in AMROC, the dimension parameter is $d=2$ or $3$ according to the dimension used, and $|\Omega|$ is the cell area.

\subsection{Clustering algorithm}
\begin{figure}[htb]
    \centering
    \newcommand*{\xMin}{0}%
\newcommand*{\xMax}{10}%
\newcommand*{\yMin}{0}%
\newcommand*{\yMax}{15}%
\begin{center}
\resizebox{8cm}{8.5cm}{%
\begin{tikzpicture}
    [
        box/.style={rectangle,draw=black,thick, minimum size=1cm},
    ]
\path (14,2) node (idn5) [ blue!90!black!90] {\textbf{$\Delta = \Upsilon_{k-1} - 2 \Upsilon_{k} + \Upsilon_{k+1} $}};

\foreach \x in {1,...,10}{
    \foreach \y in {1,...,10}
        \node[box,gray!55] at (\x,\y){};
}

\foreach \x in {4,...,10} {
\node[box,fill=black!10!green!20] at (\x,9){};
\node[box,fill=black!10!green!20] at (\x,10){};
 }
\foreach \x in {3,...,5} {
\node[box,fill=red!15] at (\x,8){};
 }
\node[box,fill=red!15] at (3,7){};

\foreach \x in {1,...,2} {
\node[box,fill=blue!5] at (\x,7){};
 }
\foreach \x in {1,...,2} {
\node[box,fill=red!5] at (\x,7){};
\node[box,fill=red!5] at (\x,6){};
\node[box,fill=red!5] at (\x,5){};
\node[box,fill=red!5] at (\x,4){};
\node[box,fill=blue!20] at (\x,2){};
\node[box,fill=blue!20] at (\x,1){};
 }
\path (11,10.7) node (idn5) [ ] {{$\Upsilon^{\textcolor{blue}{1}}_{\mbox{row}}$}};
\path (11,10) node (idn5) [ ] {{$6$}}; 
\path (11,9) node (idn5) [ ] {{$6$}}; 
\path (11,8) node (idn5) [ ] {{$3$}}; 
\path (11,7) node (idn5) [ ] {{$3$}}; 
\path (11,6) node (idn5) [ ] {{$2$}}; 
\path (11,5) node (idn5) [ ] {{$2$}}; 
\path (11,4) node (idn5) [ ] {{$2$}}; 
\path (11,3) node (idn5) [ ] {{\textcolor{blue}{$0$}}}; 
\path (11,2) node (idn5) [ ] {{$2$}};
\path (11,1) node (idn5) [ ] {{$2$}};
\path (12.3,10.7) node (idn5) [ ] {{$\Delta^{\textcolor{blue}{1}}_{\mbox{row}}$}};
\path (12,9) node (idn5) [ ] {\textcolor{blue}{$-3$}}; 
\path (12,8) node (idn5) [ ] {\textcolor{blue}{$\;\;3$}}; 
\path (12,7) node (idn5) [ ] {{$-1$}}; 
\path (12,6) node (idn5) [ ] {{$\;\;1$}}; 
\path (12,5) node (idn5) [ ] {{$\;\;0$}};

\path (13.5,9) node (idn5) [ ] {{$\Upsilon^{\textcolor{red}{2}}_{\mbox{row}}$}};
\path (13.5,8) node (idn5) [ ] {{$3$}}; 
\path (13.5,7) node (idn5) [ ] {{$3$}}; 
\path (13.5,6) node (idn5) [ ] {{$2$}}; 
\path (13.5,5) node (idn5) [ ] {{$2$}}; 
\path (13.5,4) node (idn5) [ ] {{$2$}}; 
\path (14.7,9) node (idn5) [ ] {{$\Delta^{\textcolor{red}{2}}_{\mbox{row}}$}};
\path (14.5,7) node (idn5) [ ] {\textcolor{red}{$-1$}}; 
\path (14.5,6) node (idn5) [ ] {\textcolor{red}{$\;\;1$}}; 
\path (14.5,5) node (idn5) [ ] {{$0$}};

\path (15.8,9) node (idn5) [ ] {{$\Upsilon^{\textcolor{brown}{3}}_{\mbox{row}}$}};
\path (15.7,8) node (idn5) [ ] {{$3$}}; 
\path (15.7,7) node (idn5) [ ] {{$1$}}; 

\path (0.2,0) node (idn5) [ ] {{$\Upsilon^{\textcolor{blue}{1}}_{\mbox{col}}$}};
\path (1,0) node (idn5) [ ] {{$6$}}; 
\path (2,0) node (idn5) [ ] {{$6$}}; 
\path (3,0) node (idn5) [ ] {{$2$}}; 
\path (4,0) node (idn5) [ ] {{$3$}}; 
\path (5,0) node (idn5) [ ] {{$3$}}; 
\path (6,0) node (idn5) [ ] {{$2$}}; 
\path (7,0) node (idn5) [ ] {{$2$}}; 
\path (8,0) node (idn5) [ ] {{$2$}}; 
\path (9,0) node (idn5) [ ] {{$2$}};
\path (10,0) node (idn5) [ ] {{$2$}};
\path (0.2,-1) node (idn5) [ ] {{$\Delta^{\textcolor{blue}{1}}_{\mbox{col}}$}};
\path (2,-1) node (idn5) [ ] {\textcolor{blue}{{$-4$}}}; 
\path (3,-1) node (idn5) [ ] {\textcolor{blue}{{$\;5$}}}; 
\path (4,-1) node (idn5) [ ] {{$1$}}; 
\path (5,-1) node (idn5) [ ] {{$-1$}}; 
\path (6,-1) node (idn5) [ ] {{$1$}}; 
\path (7,-1) node (idn5) [ ] {{$0$}}; 
\path (8,-1) node (idn5) [ ] {{$0$}}; 
\path (9,-1) node (idn5) [ ] {{$0$}};
\path (0.2,-2) node (idn5) [ ] {{$\Upsilon^{\textcolor{blue}{2}}_{\mbox{col}}$}};
\path (1,-2) node (idn5) [ ] {{$2$}}; 
\path (2,-2) node (idn5) [ ] {{$2$}}; 
\path (3,-2) node (idn5) [ ] {\textcolor{blue}{$0$}}; 
\path (4,-2) node (idn5) [ ] {\textcolor{blue}{$0$}}; 
\path (5,-2) node (idn5) [ ]{\textcolor{blue}{$0$}}; 
\path (6,-2) node (idn5) [ ] {\textcolor{blue}{$0$}};  
\path (7,-2) node (idn5) [ ] {\textcolor{blue}{$0$}}; 
\path (8,-2) node (idn5) [ ] {\textcolor{blue}{$0$}};  
\path (9,-2) node (idn5) [ ] {\textcolor{blue}{$0$}}; 
\path (10,-2) node (idn5) [ ] {\textcolor{blue}{$0$}}; 
\path (0,11) node (idn5) [ ] {{$\Upsilon^{\textcolor{black!30!green!90}{3}}_{\mbox{col}}$}};
\path (1,11) node (idn5) [ ] {\textcolor{black!30!green!90}{0}}; 
\path (2,11) node (idn5) [ ] {\textcolor{black!30!green!90}{0}};
\path (3,11) node (idn5) [ ] {\textcolor{black!30!green!90}{0}};
\path (4,11) node (idn5) [ ] {{$2$}}; 
\path (5,11) node (idn5) [ ] {{$2$}}; 
\path (6,11) node (idn5) [ ] {{$2$}}; 
\path (7,11) node (idn5) [ ] {{$2$}}; 
\path (8,11) node (idn5) [ ] {{$2$}}; 
\path (9,11) node (idn5) [ ] {{$2$}};
\path (10,11) node (idn5) [ ] {{$2$}};
\path (0,3.3) node (idn5) [ ] {{$\Upsilon^{\textcolor{red}{4}}_{\mbox{col}}$}};
\path (1,3.3) node (idn5) [ ] {{$4$}}; 
\path (2,3.3) node (idn5) [ ] {{$4$}}; 
\path (3,3.3) node (idn5) [ ] {{$2$}}; 
\path (4,3.3) node (idn5) [ ] {{$3$}}; 
\path (5,3.3) node (idn5) [ ] {{$3$}}; 
\path (6,3.3) node (idn5) [ ] {{$2$}}; 
\path (7,3.3) node (idn5) [ ] {{$2$}}; 
\path (8,3.3) node (idn5) [ ] {{$2$}}; 
\path (9,3.3) node (idn5) [ ] {{$2$}};
\path (10,3.3) node (idn5) [ ] {{$2$}};

\path (0,2.6) node (idn5) [ ] {{$\Delta^{\textcolor{red}{4}}_{\mbox{col}}$}};
\path (2,2.7) node (idn5) [ ] {\textcolor{red}{{$-2$}}}; 
\path (3,2.7) node (idn5) [ ] {\textcolor{red}{{$\;3$}}}; 
\path (4,2.7) node (idn5) [ ] {\textcolor{red}{{$-2$}}}; 
\path (5,2.7) node (idn5) [ ] {{$-1$}}; 
\path (6,2.7) node (idn5) [ ] {{$1$}}; 
\path (7,2.7) node (idn5) [ ] {{$0$}}; 
\path (8,2.7) node (idn5) [ ] {{$0$}}; 
\path (9,2.7) node (idn5) [ ] {{$0$}};

\path (0,12.5) node (idn5) [ ] {{$\Upsilon^{\textcolor{orange}{5}}_{\mbox{col}}$}};
\path (1,12.5) node (idn5) [ ] {{$4$}}; 
\path (2,12.5) node (idn5) [ ] {{$4$}}; 
\path (3,12.5) node (idn5) [ ] {{$2$}}; 
\path (4,12.5) node (idn5) [ ] {{$1$}}; 
\path (5,12.5) node (idn5) [ ] {{$1$}}; 
\path (6,12.5) node (idn5) [ ] {\textcolor{orange}{{$0$}}};
\path (7,12.5) node (idn5) [ ] {\textcolor{orange}{{$0$}}};
\path (8,12.5) node (idn5) [ ] {\textcolor{orange}{{$0$}}};
\path (9,12.5) node (idn5) [ ] {\textcolor{orange}{{$0$}}};
\path (10,12.5) node (idn5) [ ] {\textcolor{orange}{{$0$}}};
\path (0,11.7) node (idn5) [ ] {{$\Delta^{\textcolor{orange}{5}}_{\mbox{col}}$}};
\path (2,11.7) node (idn5) [ ] {\textcolor{orange}{{$-2$}}}; 
\path (3,11.7) node (idn5) [ ] {\textcolor{orange}{{$1$}}};
\path (4,11.7) node (idn5) [ ] {{$1$}};

\path (0,14.3) node (idn5) [ ] {{$\Upsilon^{\textcolor{brown}{6}}_{\mbox{col}}$}};
\path (3,14.3) node (idn5) [ ] {{$2$}}; 
\path (4,14.3) node (idn5) [ ] {{$1$}}; 
\path (5,14.3) node (idn5) [ ] {{$1$}}; 
\path (0,13.5) node (idn5) [ ] {{$\Delta^{\textcolor{brown}{6}}_{\mbox{col}}$}};
\path (4,13.5) node (idn5) [ ] {$1$};

\draw[very thick,dashed,blue] (3,0.5) -- (3,2.5);
\draw[very thick,dashed,blue] (0.5,3) -- (10.5,3);
\draw[very thick,dotted,blue] (0.5,8.5) -- (10.5,8.5);
\draw[very thick,dashed,green] (3,10.5) -- (3,8.5);
\draw[very thick,dashed,orange] (5.5,8.5) -- (5.5,3.5);
\draw[very thick,dotted,orange] (2.5,8.5) -- (2.5,3.5);
\draw[very thick,dotted,red] (3.5,8.5) -- (3.5,3.5);
\draw[very thick,dotted,red] (3.5,6.5) -- (10.5,6.5);

\end{tikzpicture}
}
\end{center}
    \caption{\label{fig:clusterScheme}
    Illustration of the clustering algorithm in three steps for rows and columns, with the respective notation 
    $
    \Upsilon^{\mbox{\,step}}_{\mbox{row/col}}
    $\, and $
    \Delta^{\mbox{\,step}}_{\mbox{row/col}}
    $\,.}
\end{figure}
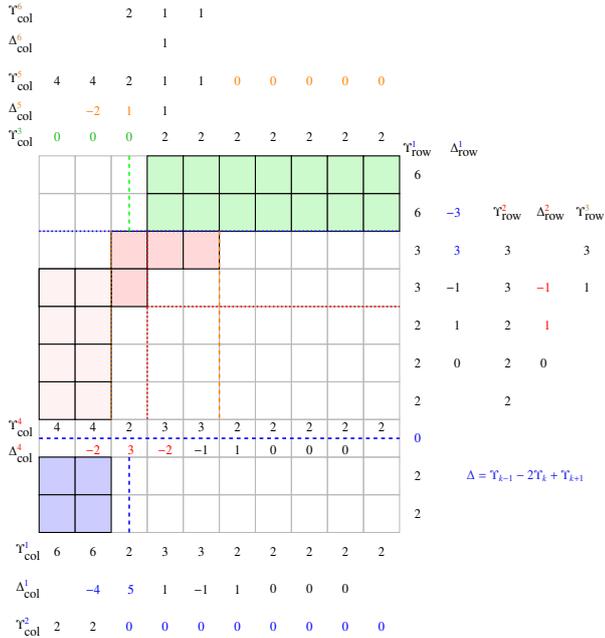

After evaluating the
refinement indicators and flagging cells for refinement, a special clustering algorithm 
\cite{Bell-Berger-Saltzman-94} is used to create new refinement patches until the ratio between all cells and flagged ones in every new submesh is above a prescribed value $0<\eta \le 1$.
Central to the \textcolor{black}{patch}-based mesh refinement approach is the utilisation of a dedicated algorithm to create blocks (or patches) from individual cells tagged for refinement by any of the above criteria.
We use a recursive algorithm proposed by Bell {\it et al.}~\cite{Belletall:1994}. This method, inspired by techniques used in image detection, counts the number of flagged cells in each row
and column on the entire domain. The sums $\Upsilon$ are called {\it signatures}. 
First, cuts into new boxes are placed on all edges where $\Upsilon$ is equal to zero. In the second step, cuts are placed at zero 
crossings of the discrete second derivative $\Delta = \Upsilon_{\nu+1}-2\,\Upsilon_\nu+\Upsilon_{\nu-1}$. The algorithm starts with the steepest zero crossing and uses recursively weaker ones, until the ratio between all cells and flagged ones in every new mesh are above the prescribed value $\eta$. 
An illustration of the general clustering procedure is given in Figure~\ref{fig:clusterScheme}.

\section{Computational experiments and discussions}
\label{secExperiments}

In order to compare and assess different refinement criteria quantitatively, we compute the $L_1$ error 
between the adaptive solution and its corresponding uniform mesh solution related to the same maximal resolution used in the adaptive computation, as discussed in 
\cite{Deiterdingetal2016SIAM}. 
We denote this error by $L_{1,\mathrm{AMR}}$.

Computations are run in parallel on nodes of a recent GNU/LINUX compute cluster that provides $20$ cores with shared memory per node. The AMROC system is parallelised through the MPI library with dynamic re-partitioning. Load balancing is carried out after each level-0 time step in the adaptive cases. AMROC is pursuing a rigorous domain decomposition strategy, in which the increased computational expense on higher refinement levels in the \textcolor{black}{patch}-based AMR algorithm is considered in evaluating parallel workload; however, only units of smallest resolution corresponding to a cell on level zero are utilised \cite{Deiterding-05}. This approach simplifies the implementation and reduces the expense of the partitioning algorithm, but it can lead to slight load imbalances on deep hierarchies. The algorithm used for partitioning is always a multi-dimensional space filling curve \cite{Deiterding-PhDThesis,DeiterdingESAIM:2011}. 

In all experiments we use a Cartesian mesh  with HLLD numerical flux introduced in \cite{MiyoshiKusano:2005} and a MinMod limiter as discussed in~\cite{Toro:1999}.

\subsection{Orszag--Tang vortex}

Nowadays, this classical experiment is also used to test how robust a compressible MHD code is at handling the formation of MHD shocks \cite{LandauLifshitz:2004},  shock-shock interactions, discontinuities, and other structure formations, as presented  in \cite{Zacharyetal:1994}.
Therefore,  this test is also frequently used for code verification and comparison, and it can also demonstrate how significant magnetic monopoles,  \textit{i.e} the $\nabla \cdot B =0$ condition, affect the numerical solution, as discussed in \cite{londrillo2000high} and references therein. 
Here we use this test to verify how the wavelet-based criteria handle shock formations.
We also present a comparison with the Scaled Gradient (SG) refinement criteria already implemented in the AMROC framework \cite{DeiterdingESAIM:2011,Deiterdingetal2016SIAM,DEITERDING:2019CAFMRSG}.
For SG criteria in two-dimensions, a cell at a position $(j,k)$ is flagged for refinement if at least one of the following relations is satisfied for  mass density, for instance, 
\[
|\,\rho_{j+1,\, k} - \rho_{j,\, k} \,|  > \epsilon^{\,\rho}, \,
|\,\rho_{j,\, k+1} - \rho_{j,\, k}\,|   > \epsilon^{\,\rho},\, 
|\,\rho_{j+1,\, k+1} - \rho_{j,\, k}\,| > \epsilon^{\,\rho},\,
\]
where the constant $\epsilon^{\,\rho}$ denotes the prescribed refinement limit; 
in subsequent parts we simply denote it by $\epsilon=\epsilon^{\,\rho}$.

\subsubsection*{Computational set-up} 
We consider as initial conditions the density $\rho= \gamma^2$,  the pressure $p=\gamma$, and the periodic velocities  with $u_x = - \sin(y)$, and $u_y = \sin(x)$ together with the magnetic field $B_x=-\sin(y)$, $B_y=\sin(2x)$, using the parameters $\gamma = 5/3$ and CFL $\nu=0.3$. 
The computational domain is $0~\leq~x~\leq~1$, $0~\leq~y~\leq~1$ with periodic boundary conditions. 
This magnetic field is constructed using a periodic vector potential to guarantee vanishing divergence of the magnetic field.

\subsubsection*{Numerical results}
\begin{figure}[htb]
    \centering
\begin{tabular}{cc}
(a) Uniform mesh  &\\ 
\includegraphics[width=0.35\columnwidth]{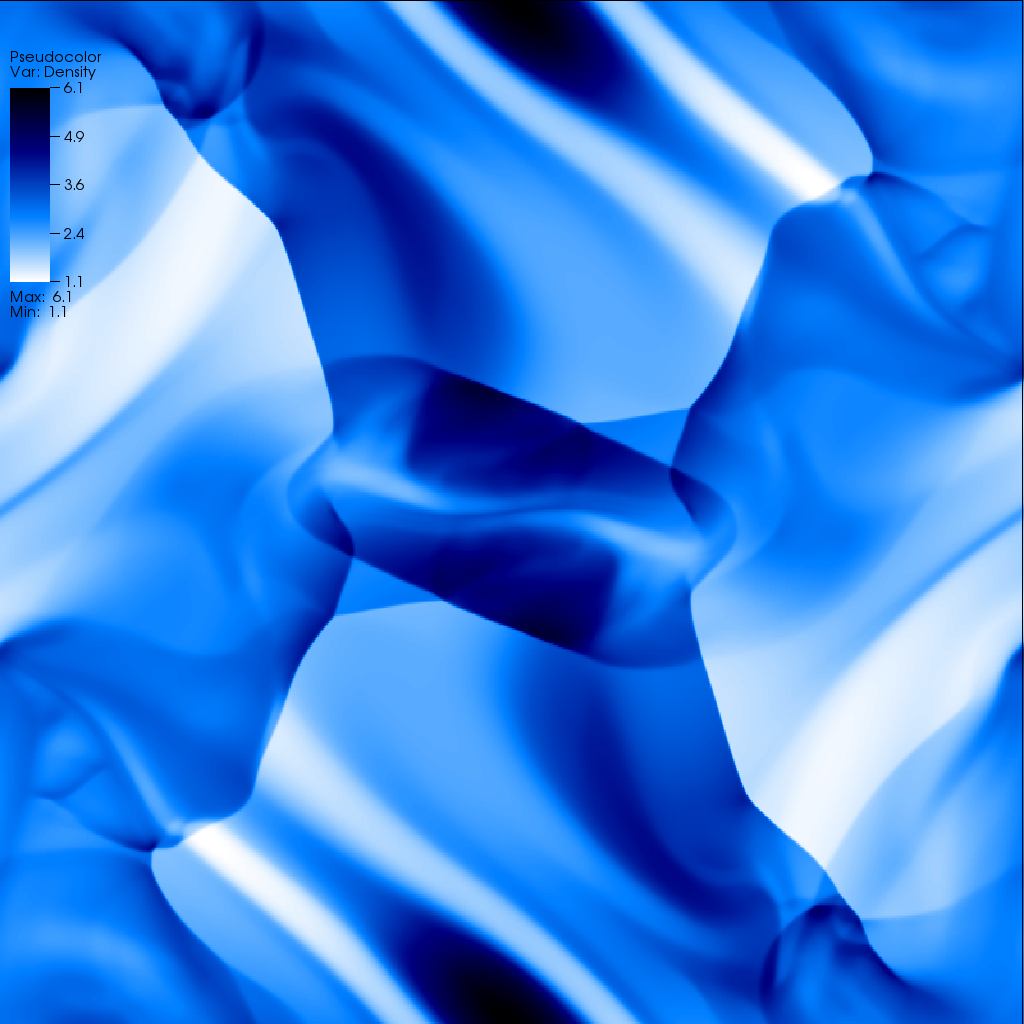} &
\includegraphics[width=0.35\columnwidth]{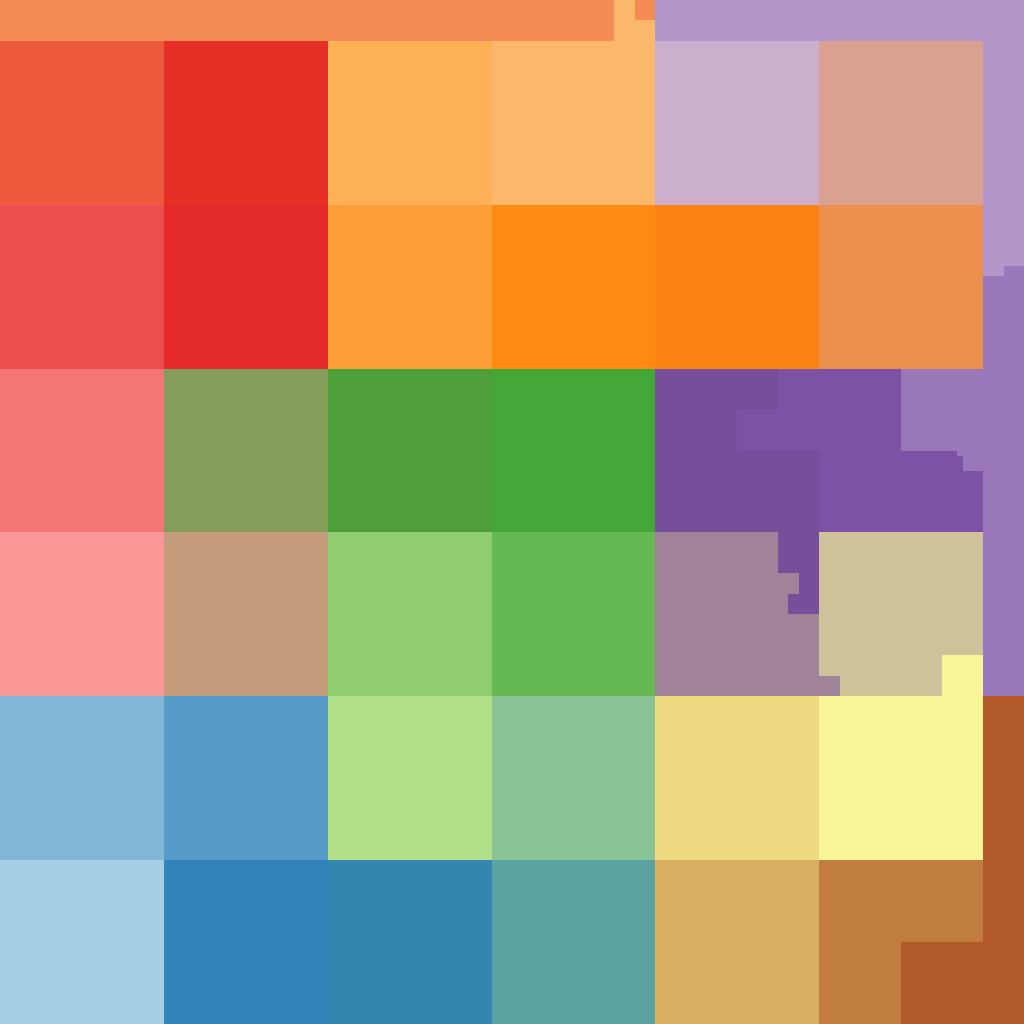}\\
(b) SG  \\
\includegraphics[width=0.35\columnwidth]{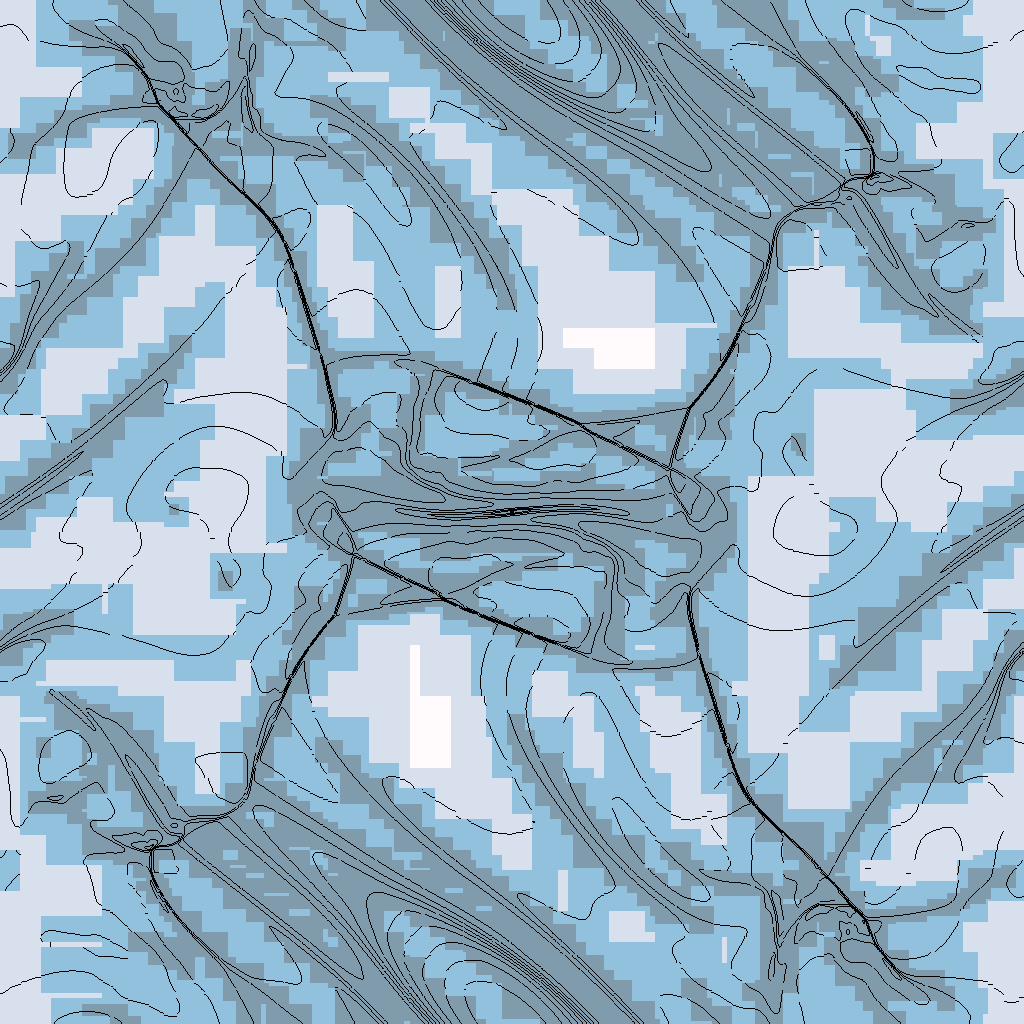} &
\includegraphics[width=0.35\columnwidth]{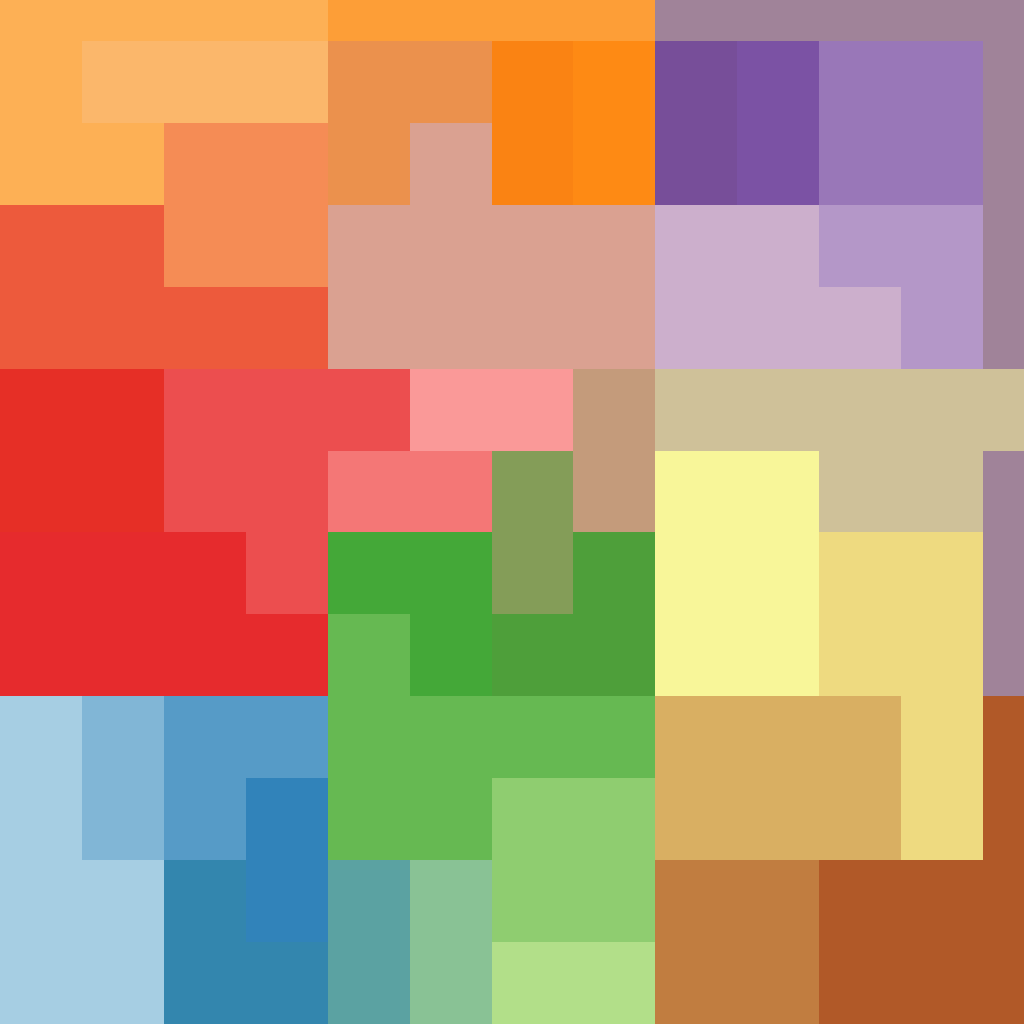} \\
\\
 (c) MR  & \\
\includegraphics[width=0.35\columnwidth]{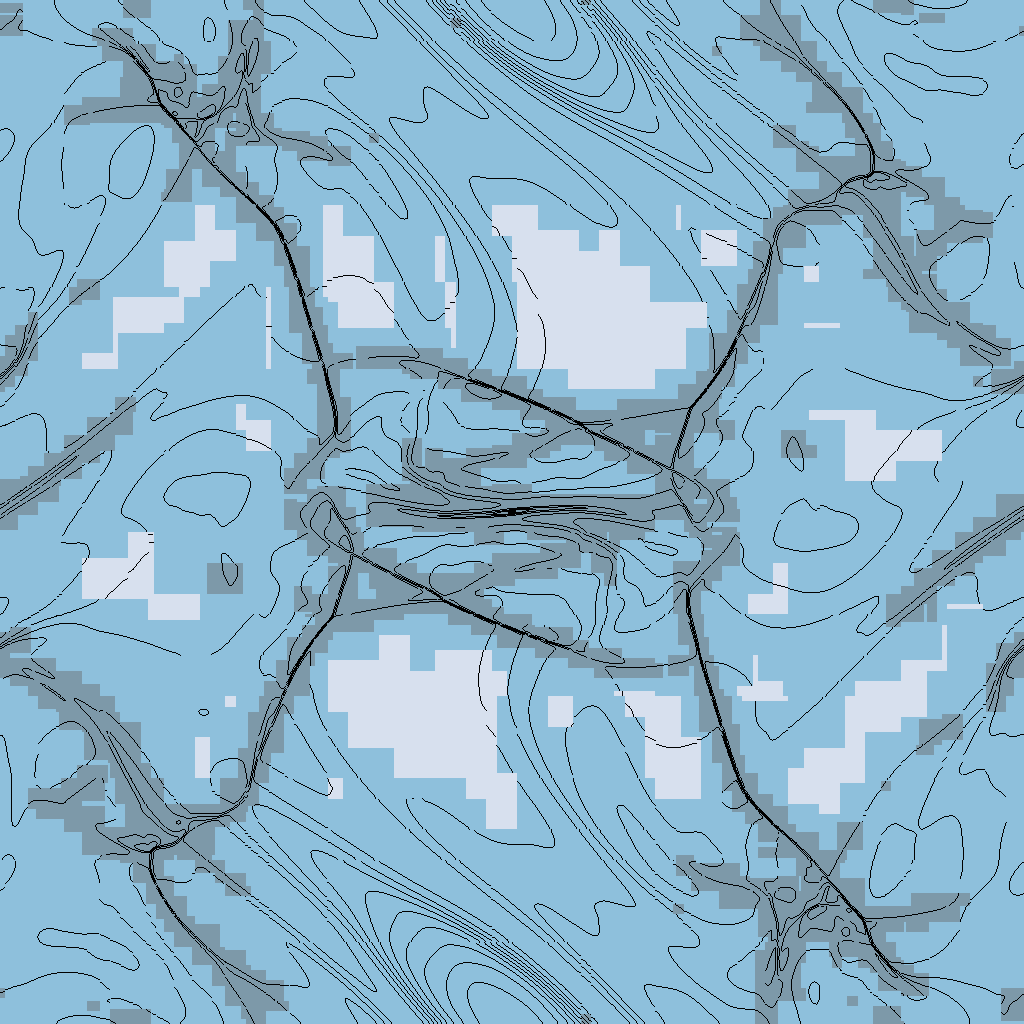} &
\includegraphics[width=0.35\columnwidth]{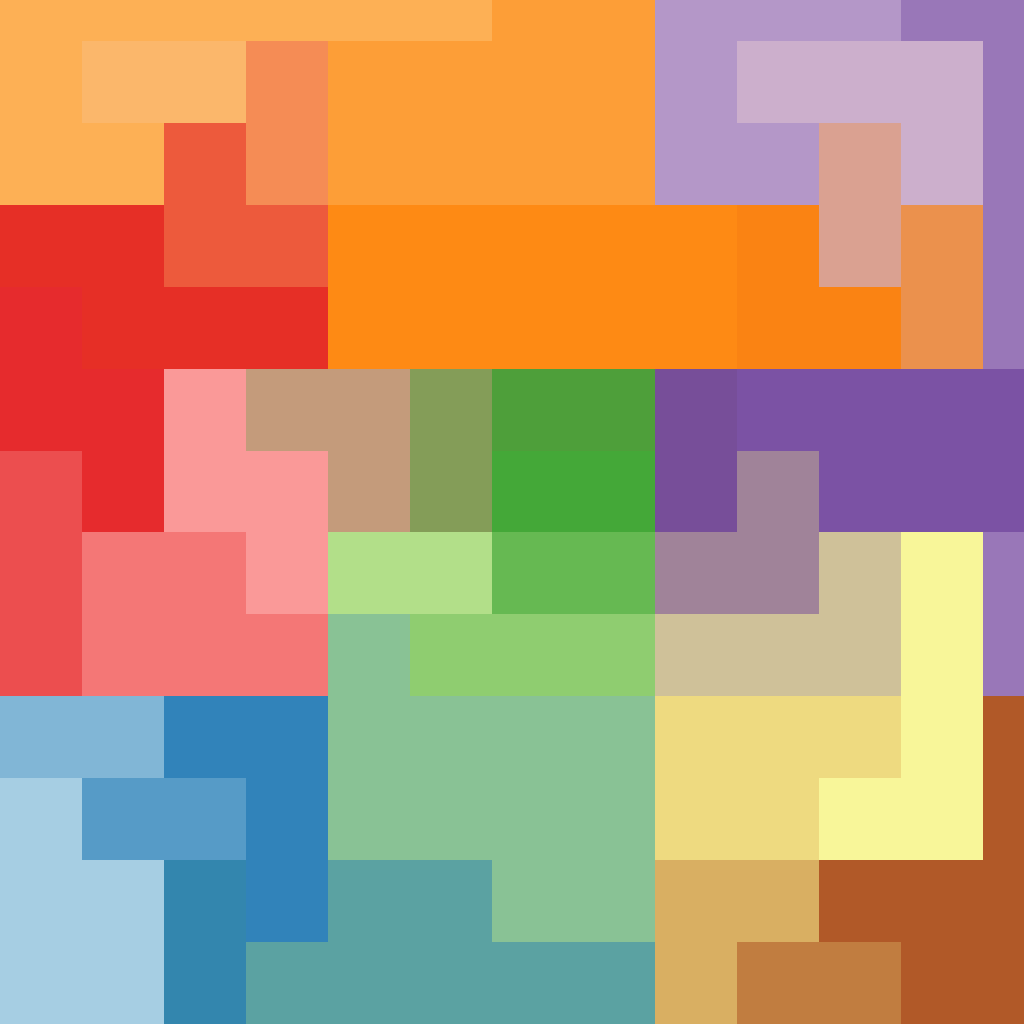}
\end{tabular}
\caption{\label{fig:OTrho} Orszag--Tang vortex: mass density at final time $t_e=\pi$. (a) uniform mesh with max value $6.16$, in black, and min value $1.1$ in white, (b, and  c)  adaptive meshes (pseudo-colour: white coarsest level $2$, grey finest level $5$) using SG, and MR criteria, respectively. 
The panels in the right column represent the data distribution in the $40$ processors indicated by colours.
}
\end{figure}

\begin{figure}[htb]
\begin{center}
\includegraphics[angle=-90,width=0.98\columnwidth]{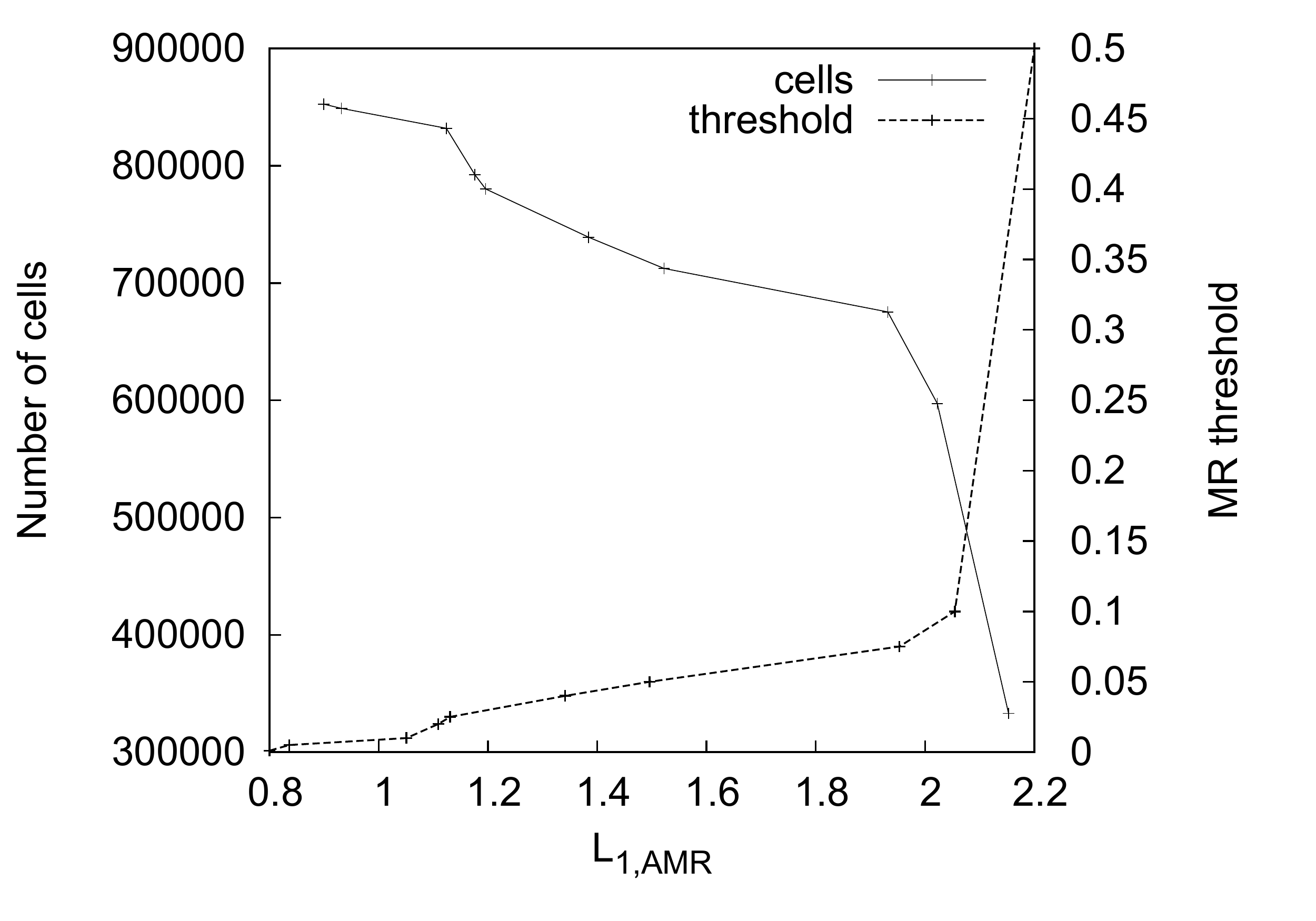}
\end{center}
    \caption{\label{fig:epscellsOT}Orszag--Tang vortex: flag cells and MR threshold at $t_e=\pi$ as a function of the $L_{1, \mathrm{AMR}}$ error of the mass density.\label{fig:OTerror} }
\end{figure}

In this experiment we have observed that the wavelet-based MR criteria detect the shock formations reproducing the expected structures.
Moreover, we have found that it produces less refinement of additional features at the maximal level compared to the gradient criterion and additionally  stronger coarsening.
Therefore, MR prevents unnecessary over-refinement while preventing improper coarsening. 
These representative effects are the reason why the MR criterion with hierarchical thresholding achieves a smaller error than the SG criterion.
An example of these effects is presented in Figure~\ref{fig:OTrho} (b, and c, in the right panel).
The adaptive mass density solution is presented in Panel (a) for the uniform mesh and it is represented by isolines. 
Pseudo-colours are used to identify the levels. 
These results are computed for $L=5$ maximum refinement level, coarsest mesh $32^2$, \textit{i.e.,} corresponding to a uniform mesh of $512^2$ cells,  at final time $t_e=\pi$.
\textcolor{black}{
To compare the number of cells we have used in Table~\ref{tab:accuracy} the value $\eta=0.80$ and threshold parameters which lead to similar errors for SG and MR.
Therefore, even with slightly smaller errors, MR needs less cells to obtain the expected representation, resulting in a gain of $1.3$ considering the ratio between the number of cells and the error.}
\begin{table}[htb]
\centering
\caption{\label{tab:accuracy}\textcolor{black}{ MR and SG: accuracy and number of cells.}}
\begin{tabular}{cccc}
\toprule
Method & $\epsilon$ & $L_{1,\mathrm{AMR}}$ & \# of cells\\
\midrule
SG & $0.20$ & $2.30$ & $411,392$ \\
MR & $0.05$ & $2.15$ & $333,064$\\
\bottomrule
\end{tabular}
\end{table}
This result is highlighted in panels (b) and (c) on the right of Figure~\ref{fig:OTrho}. 
The SG method presents refinement from levels $2\!-\!5$, whereas the MR method from $3\!-\!5$. 
We have also observed that SG uses much more refinement at the highest level.
On the left panel, we present the processor distribution for the three cases.
All show a good distribution on the processors. 
Especially for the adaptive cases, the distribution pattern follows the data organisation on the level refinements. As the MR case has less blocks in the highest level and more regions at the same level, its distribution is slightly less fragmented than for the SG one.  
\textcolor{black}{
The accuracy of the adaptive solution depends on the choice of the MR threshold parameter with a computational cost determined by the number of cells used in the representation. In our case the number of cells is directly related to the CPU time.
Figure~\ref{fig:epscellsOT} illustrates the accuracy behaviour measured by the $L_{\mathrm{1,AMR}}$ error, which is related to the choice of the threshold (vertical axis at right). 
Therefore, we can estimate roughly the error when choosing the threshold based on this behaviour.
On the left vertical axis, we present the number of cells. 
The number of cells in the adaptive mesh increases as the error decreases. Again, we can deduce a rough estimator how to proceed for choosing a certain threshold to obtain a certain number of cells based on the desired accuracy.
} 
\clearpage
\subsection{Magnetic shock cloud}

This experiment models a disruption of a high-density magnetic cloud by a strong shock wave, as described in \cite{TOUMA2006617}. 
The initial condition of this Riemann problem defines a region of the advancing plasma -- which causes the shock -- and a stationary state where the shock advances. We also define a spherical magnetic cloud as a high density region, like a plasmoid, in hydrostatic equilibrium with the surrounding plasma. 

We compare the performance obtained with the serial multiresolution Carmen-MHD code \cite{Gomes:2018:SiNuMo,Gomes2015199,DominguesetalESAIM:2013,RSTB03}, which is the prototype solver we had implemented initially during this development project. 
The detailed software design aspects of both software systems are discussed in \cite{Deiterdingetal2016SIAM}.
This experiment also provides the performance of the wavelet \textcolor{black}{patch}-based method when dealing with high speed flows. 

\subsubsection*{Computational set-up}

The cloud region has its centre at $(0.25,0.5,0.5)$ and radius $r_0 = 0.15$ with  density $\rho=10$ inside the cloud and $\rho=1$ otherwise. 
We have adopted outlet boundaries and these regions are limited by the domain boundaries and a plane parallel to the $yz$ plane at $x=0.05$.
The advancing plasma initial condition is given by $\rho = 3.86859$, $p = 167.345$, $u_x = 11.2536$, $u_y=u_z=B_x=0$, $B_y = 2.1826182$, and $B_z = -B_y$, with a stationary state given by $p=1$, $\mathbf{u}=0$, $B_y = B_z = 0.56418958$. 
The computational domain is $[0,1]^d$, where $d=2$ or $3$ respectively stand for the two- and the three-dimensional case. %
We use the following parameters: CFL $\nu=0.3$, $\gamma = 5/3$, the MR threshold is applied to the pressure variable and we run the simulations until the final time $t_e=0.06$.

\subsubsection*{Numerical results}

For two-dimensional simulations the wavelet-based adaptive solutions present similar behaviour for both Carmen-MHD and AMROC solvers. 
Both capture well all the expected relevant physical structures, especially the bow shock, and moreover the symmetry of the solution is almost perfectly preserved.

\begin{figure}[htb]
\begin{center}
\begin{tabular}{cc}
(a) Carmen-MHD code  &\\ 
\includegraphics[width=0.4\columnwidth]{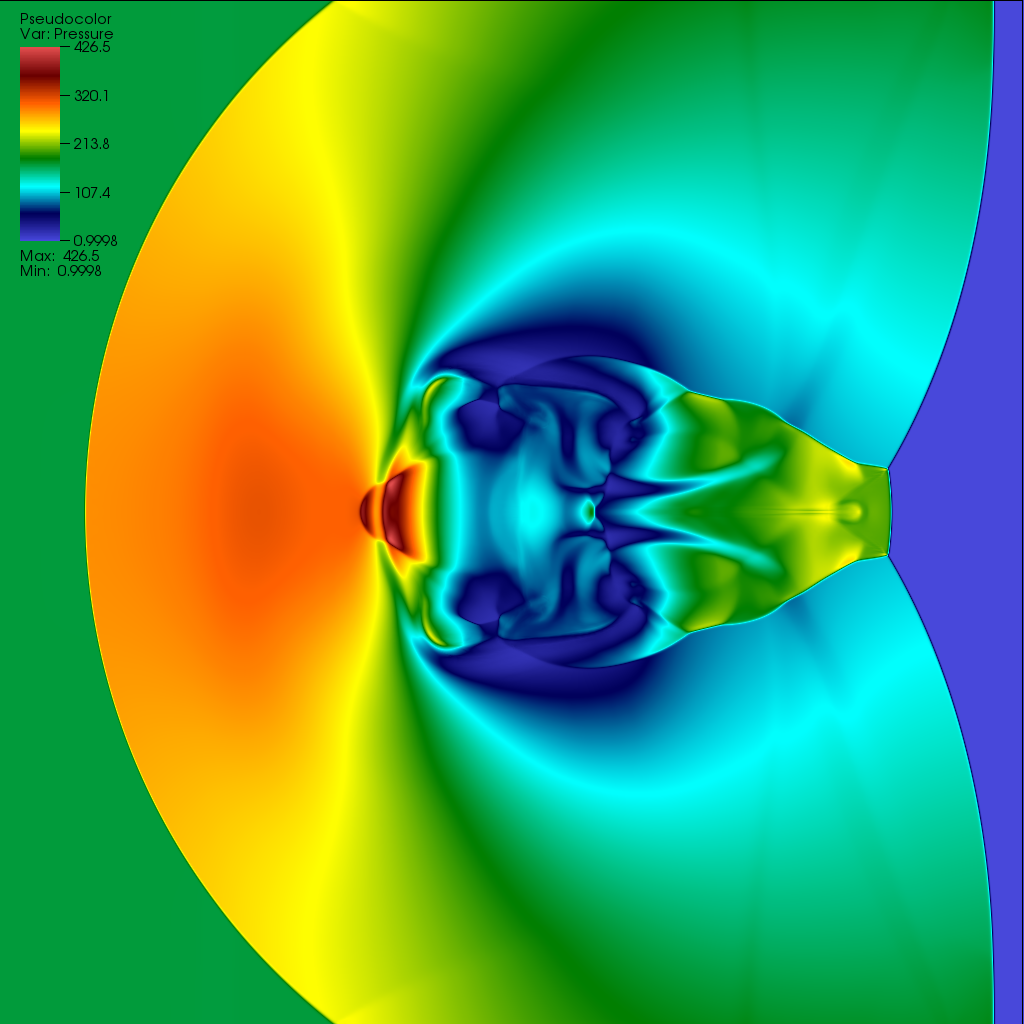} &
\includegraphics[width=0.56\columnwidth]{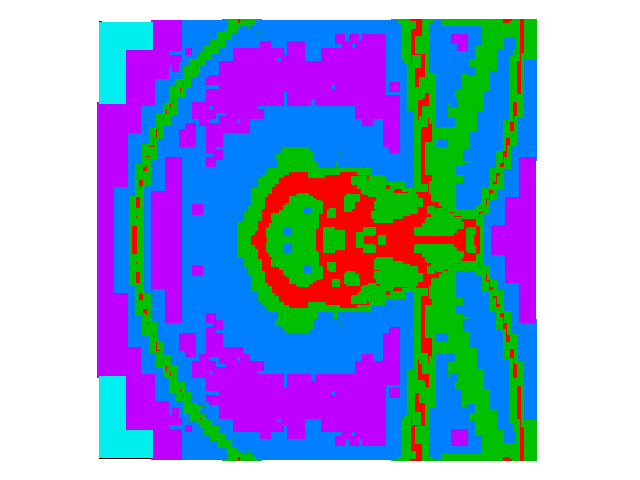}\\
(b) AMROC framework &   \\[1mm]
\includegraphics[width=0.40\columnwidth]{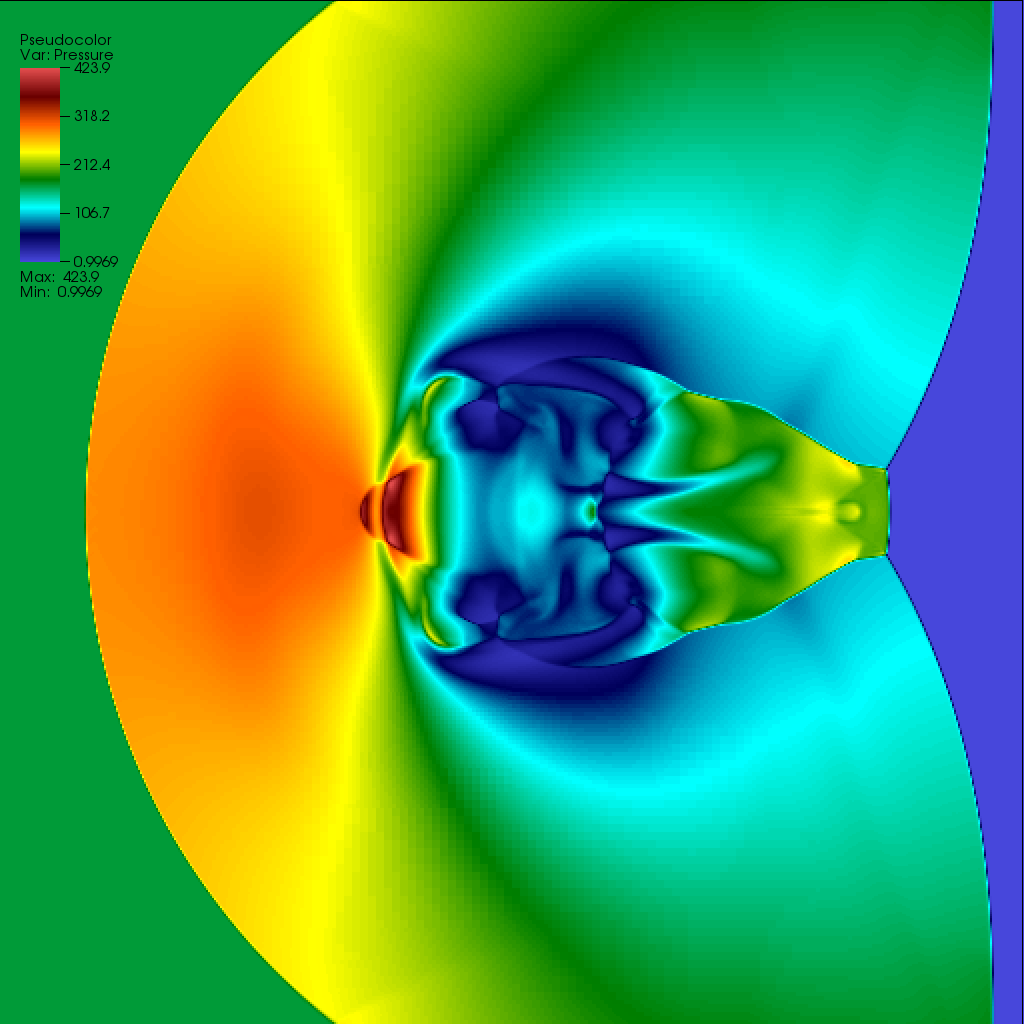} &
\includegraphics[width=0.40\columnwidth]{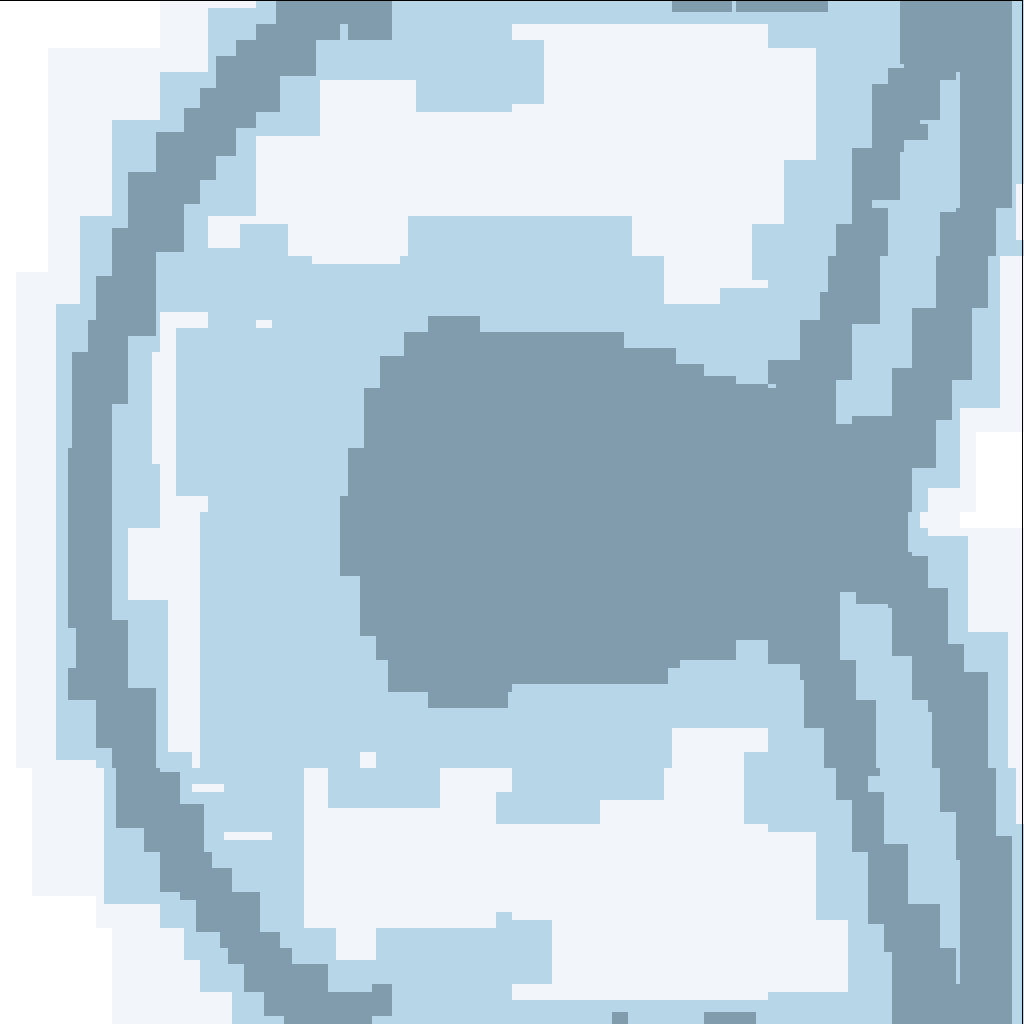} 
\end{tabular}
\end{center}
\caption{\label{fig:MC2d}  Adaptive computations of the pressure solution for the magnetic cloudy test case related to a $1,024^2$ uniform mesh. (a) Carmen-MHD code, (b) AMROC framework with $\eta=0.99$ and $16^2$ base mesh with $L=7$. 
Left columns show the adaptive meshes, red colour corresponds to the most refined level for Carmen-MHD, and dark-blue for AMROC.
}
\end{figure}

Figure~\ref{fig:MC2d} presents pressure solutions and their respective adaptive meshes at $t_e$ for both environments. 
In the Carmen-MHD code, the cell-based structure requires many less cells in the representation than in the \textcolor{black}{patch}-based one used in the AMROC framework. 
At the final time $t_e$, AMROC needs  $311,612$ cells clustered in $528$ blocks with $\eta=0.99$ and Carmen-MHD $ 40,329$ cells. 
On the other hand, we obtain an improvement in CPU time of about $9$ times in AMROC compared to Carmen-MHD, considering mono-processor runs. 
Similar results were obtained for the hydrodynamic Euler solvers in both environments \cite{Deiterdingetal2016SIAM}. 

With the additional parallelisation in the AMROC framework, the improvement becomes even larger, with a speed-up of $39$ using $16$ processors, as reflected in the CPU times in Table~\ref{tab:MC2D}.
AMROC simulations are performed with seven refinement levels over a base mesh $16^2$, using $2^n$ processors with $n$ from one to four. 
We observe that in these experiments the maximum AMROC scalability is near $1.8$ for one to two processors 
and it reduces to a factor of around $1.3$ for the four and eight processor cases. Considering eight and sixteen processors, this factor is smaller due to a reduction of the problem size, as expected.

\begin{table}[htb]
    \caption{\label{tab:MC2D} Comparison of the CPU time (s) for the two-dimensional magnetic cloudy test case corresponding to a refined mesh $1,024^2$, $\epsilon=0.01$, $\eta=0.80$, and base mesh $16^2$.}
    \label{tab:my_label}
    \begin{center}
    \begin{tabular}{lrcccccc}
    \toprule
     & \multicolumn{4}{c}{Number of Processors } \\ 
     \cmidrule{2-6}
 $2$D      & $1$   &  $2$  & $4$   &  $8$  & $16$ \\
Carmen-MHD & $17,520$  &  - & - & - & - &    \\
AMROC  & $1,889$ &   $1021$  & $635$  &  $478$ &   $447$   \\    
\bottomrule
\end{tabular}
\end{center}
\end{table}

Considering the $3$D case
with computations using one processor and a MR threshold $\epsilon=0.01$ until time $t_e$,
the Carmen-MHD code with a corresponding  $128^3$ refined mesh needs a CPU time of $16,816$ seconds, while AMROC needs only $2,620$ seconds considering a base mesh $32^2$, and $\eta=0.80$. 
Therefore, we achieve a speed-up of a factor $6$. 
Furthermore, if $60$ processors are considered, the speed-up increases to around $84$, as this AMROC computation uses only $201$ seconds, in this case with $\epsilon=0.025$ and the same $\eta=0.80$. 

\begin{figure}[htb]
\begin{center}
\begin{tabular}{p{0.45\columnwidth}p{0.45\columnwidth}}
\includegraphics[width=0.425\columnwidth]{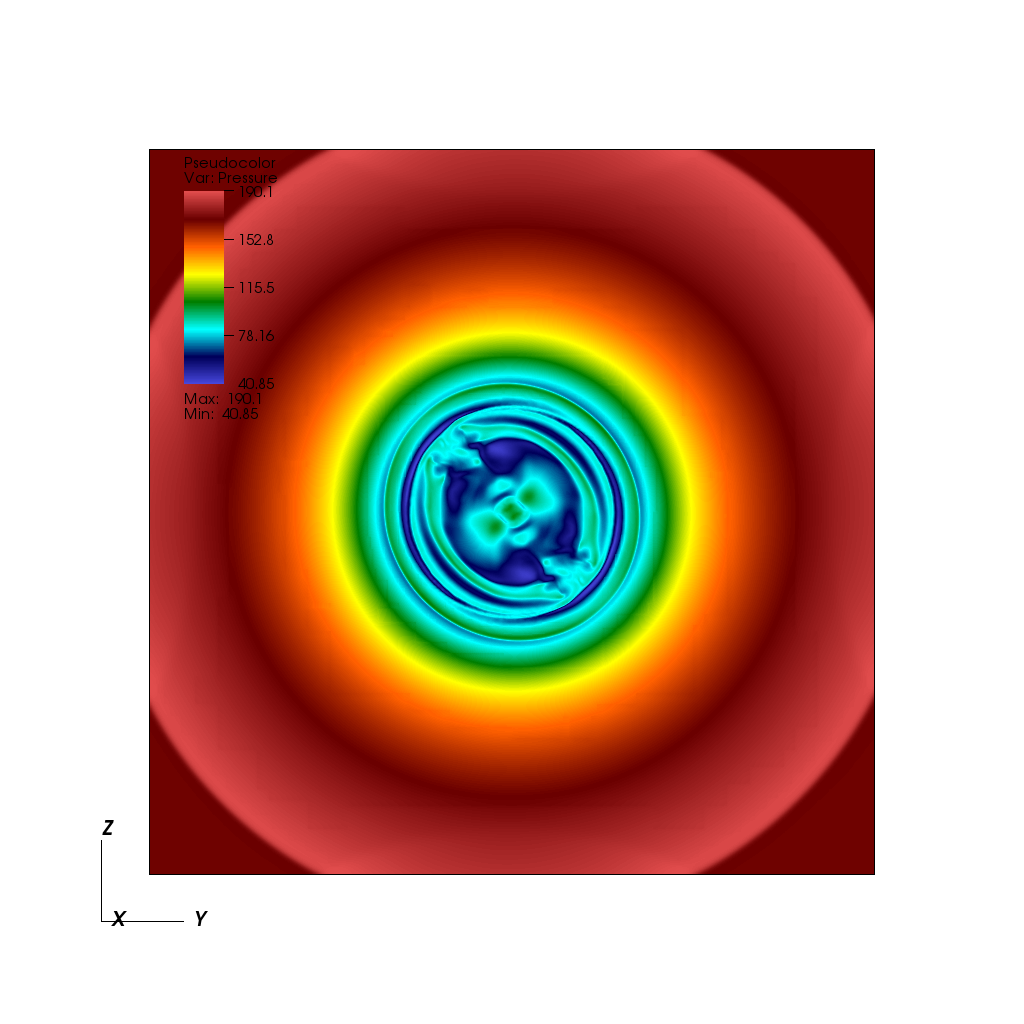}&
\includegraphics[width=0.425\columnwidth]{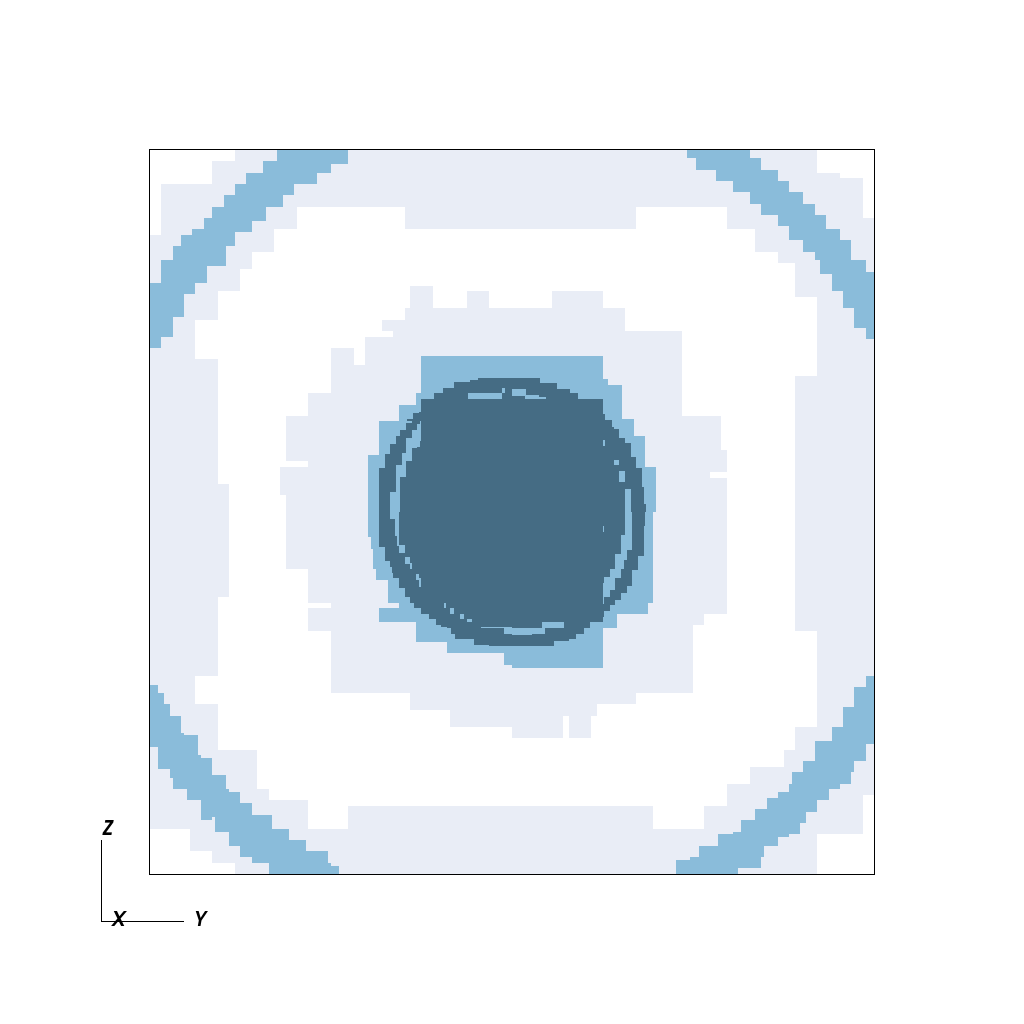}
\\[-0.5cm]
\includegraphics[width=0.425\columnwidth]{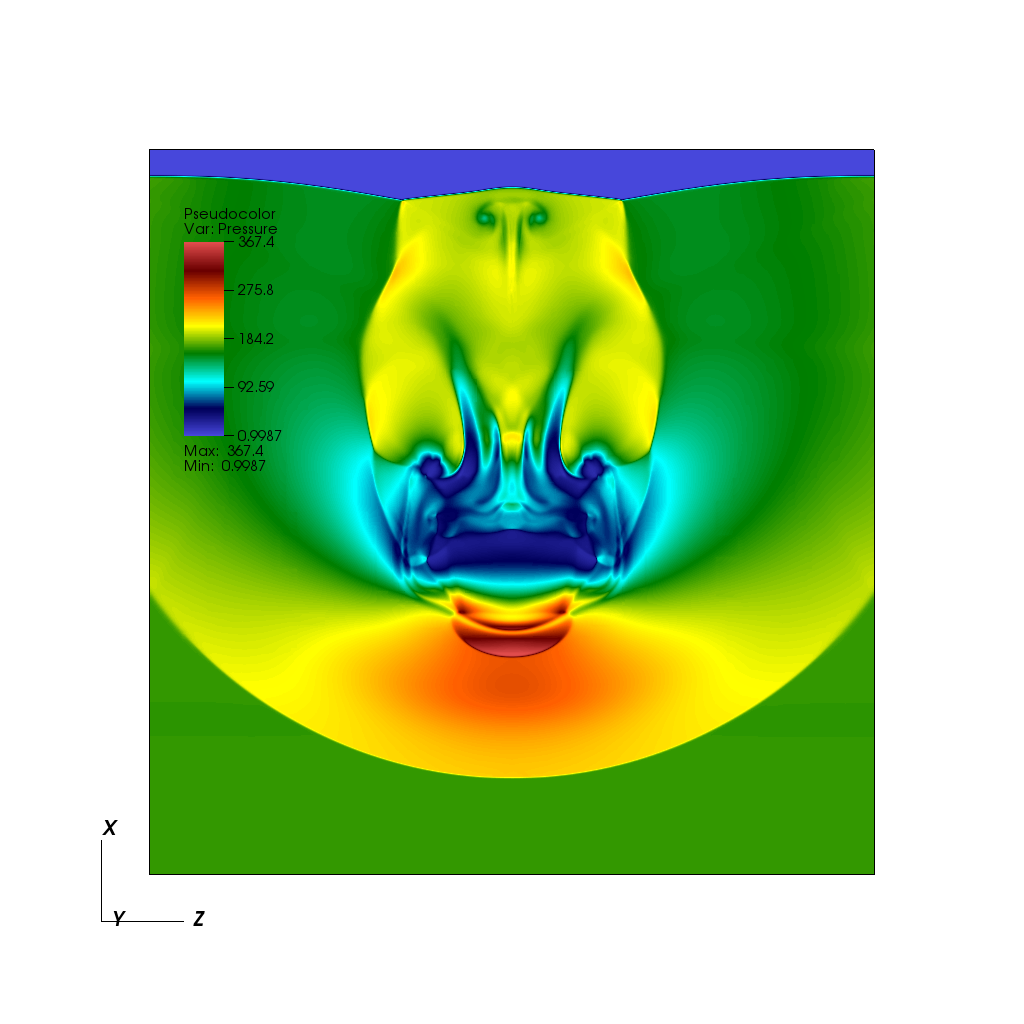}&
\includegraphics[width=0.425\columnwidth]{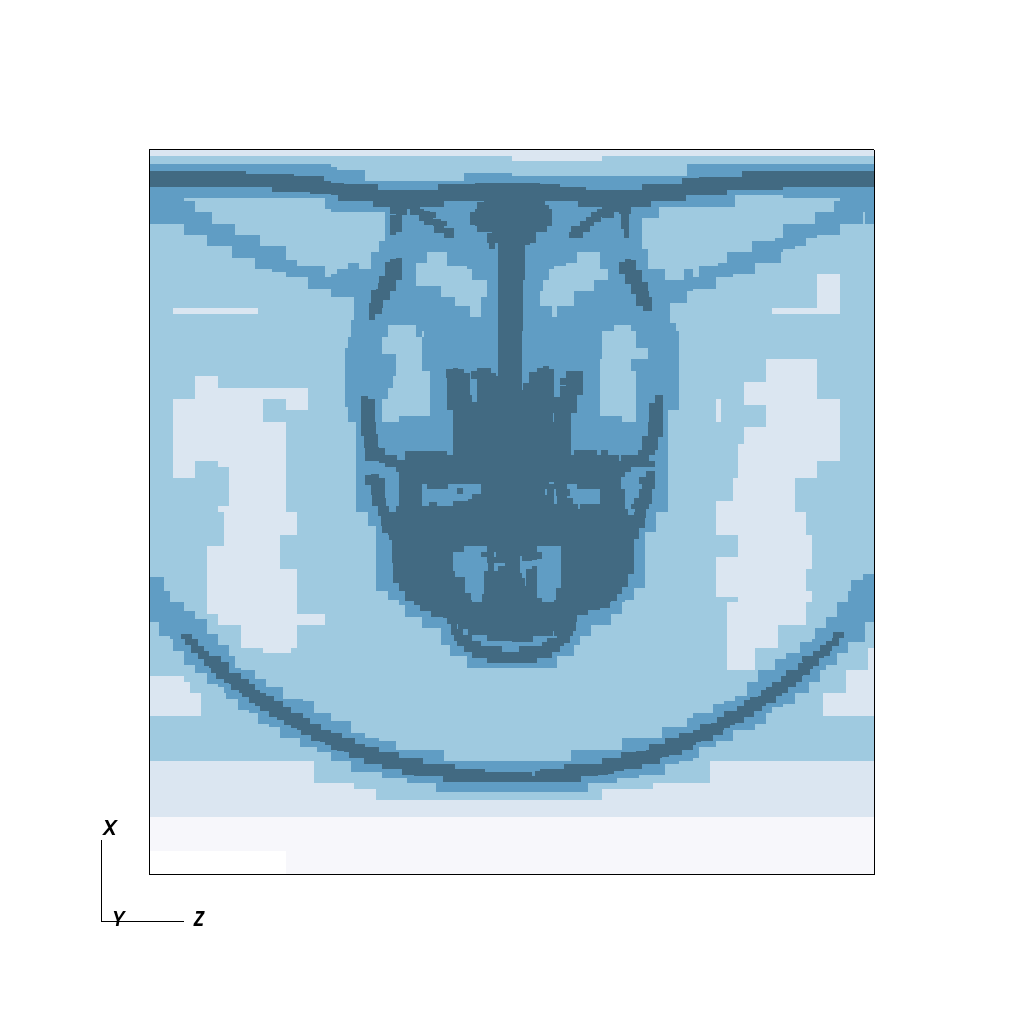}
\\[-0.5cm]
\includegraphics[width=0.425\columnwidth]{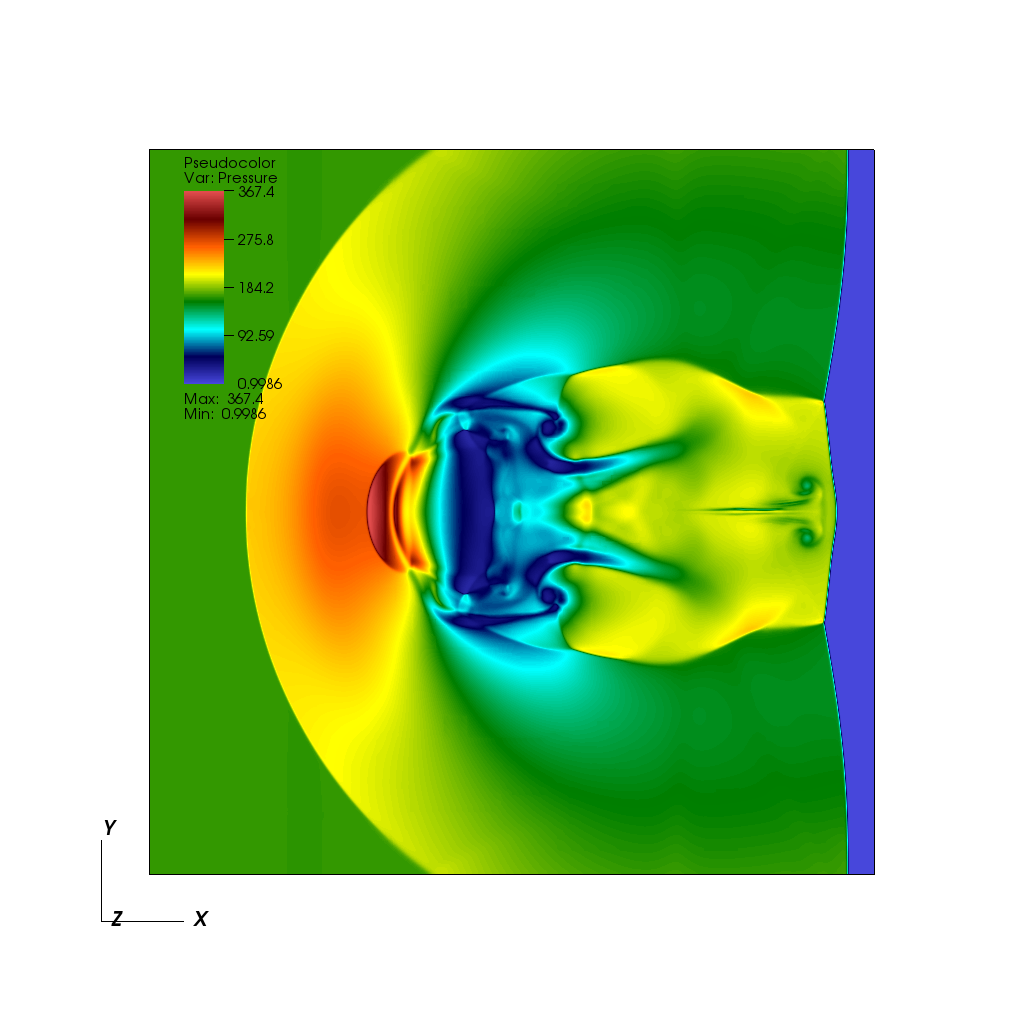}&
\includegraphics[width=0.425\columnwidth]{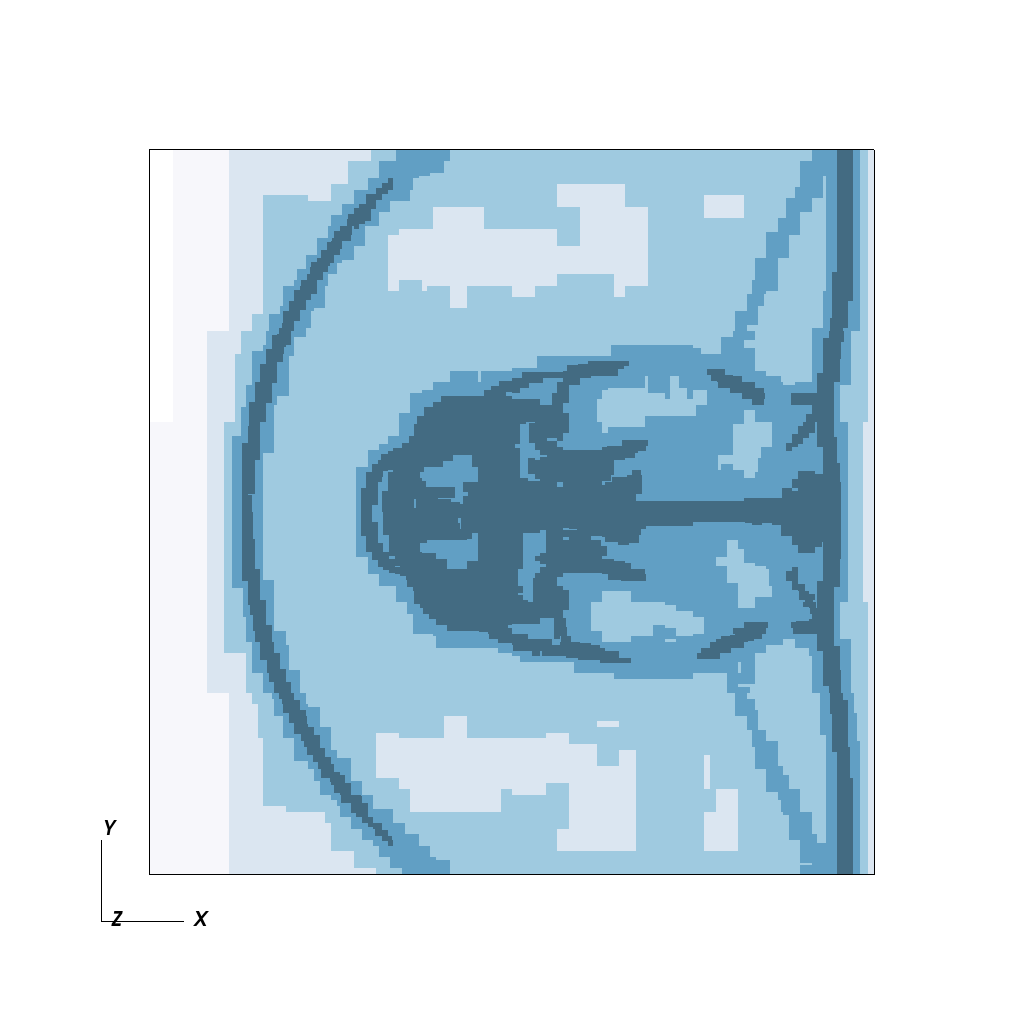}\\[-0.5cm]
\end{tabular}
\end{center}
\caption{\label{fig:MC3D} Adaptation for the 3D magnetic cloud test case at time $t_e=0.06$.  AMROC computation related to a uniform mesh $1,024^3$ with base mesh $32^3$, $6$ refinement levels,  $\epsilon=0.025$, and $\eta=0.80$. Left column: 
$2$D cuts of pressure in different planes ($y-z$ plane at $x=0.5$, $x-z$ plane with $y=0.5$, and $x-y$ plane with $z=0.5$).  Right column: Refinement levels, dark-blue is the maximum refinement level $L$.
In all panels, the orientations of the axes are according to the right-hand rule.  
}
\end{figure}

\begin{figure}[htb]
\begin{tabular}{p{0.4\columnwidth}p{0.4\columnwidth}}
\includegraphics[width=0.525\columnwidth]{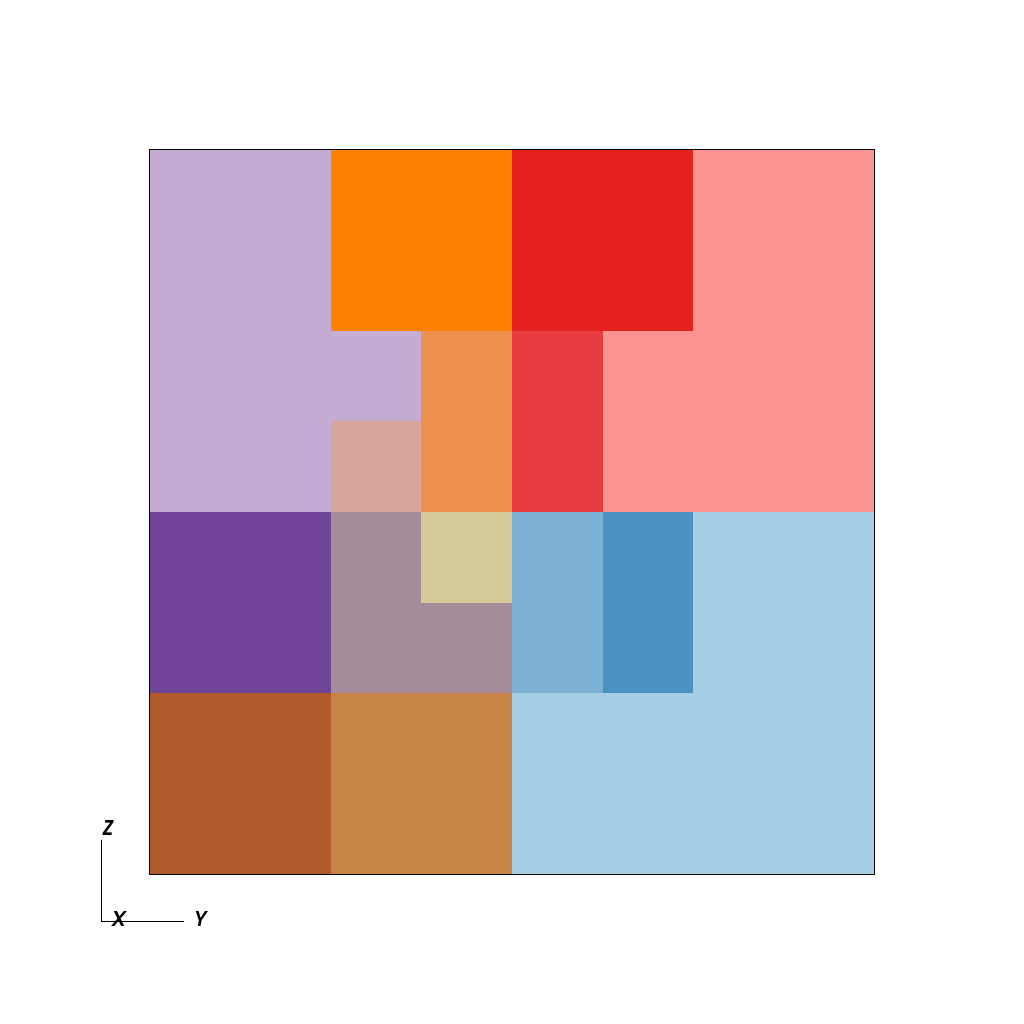} &
\includegraphics[width=0.525\columnwidth]{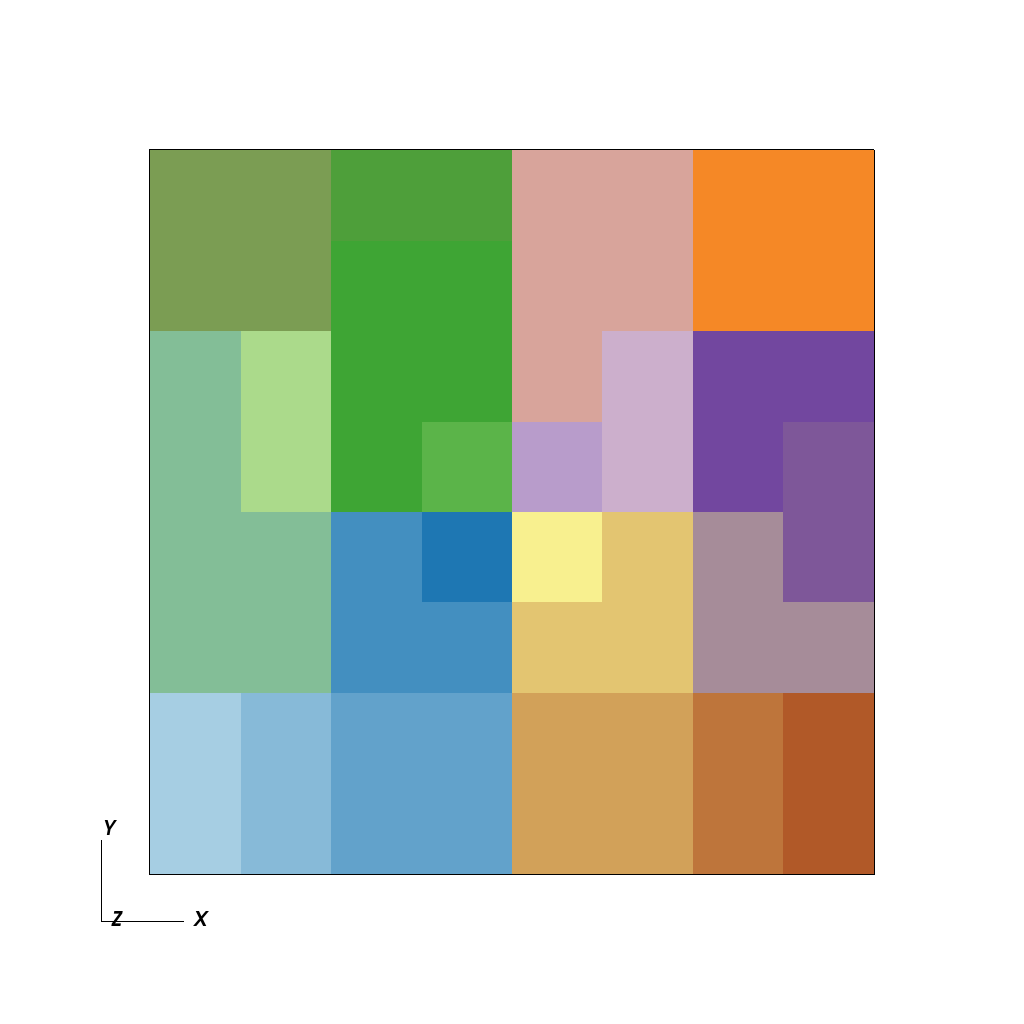}
\end{tabular}
\caption{\label{fig:MC3Ddist} Processor distribution for the $3$D magnetic cloud test case at time $t_e=0.06$. AMROC computation related to a uniform mesh $1,024^3$ with $\epsilon=0.025$ and $60$ processors. 
$2$D cuts in different planes ($y-z$ plane at $x=0.5$, and $x-y$ plane with $z=0.5$).  Colours indicate the data distribution in the processors at the cuts. 
}
\end{figure}

Two-dimensional cuts of the pressure solution on the adaptive mesh at time $t_e=0.06$ are presented in Fig.~\ref{fig:MC3D} (left panel), for $\epsilon=0.025$, $\eta=0.80$, base mesh $32^2$, and corresponding fine mesh $1024^3$. 
We can again observe that the symmetry is almost perfectly preserved and the adaptive mesh refines all relevant structures, particularly the bow shock, the front cap, and the internal structures.
Additionally, new features could be developed in the three-dimensional scenario, especially, in the tail, and in front of the cloudy structure. 
Conjointly, the mesh adaptation at time $t_e=0.06$ fits very well the main features of the solution (Fig.~\ref{fig:MC3D}, right panel).
Furthermore, the adaptive mesh distribution
is well balanced considering the difference in computational
costs required by the adaptive meshes (Fig.~\ref{fig:MC3D}, right  panel), as it can be observed in the $2$D cuts in Fig.~\ref{fig:MC3Ddist}.
The processor distribution (not shown) of the $x-z$ plane presents similar results as for the $x-y$ plane.
Using Table~\ref{tab:MC3Dcpu}, we show the $3$D performance of AMROC related to three corresponding locally refined meshes, and $60$ processors (distributed on $3$ nodes) at time $t_e=0.06$.  
We observe that the adaptation rises proportionally with the mesh size as we expand the corresponding refined mesh by a factor of two. 
Basically, a factor of $10$ increases the CPU time as we expand by a factor $2^3$ the fully refined mesh. However, the effective number of cells on the adaptive mesh growths only by a factor of   $5$.


\begin{table}[htb] \caption{\label{tab:MC3Dcpu}AMROC $3$D adaptive computations of the magnetic cloudy test case corresponding to  $\epsilon=0.025$, $\eta=0.80$, and base mesh $32^2$.
}
\begin{center}
\resizebox{\columnwidth}{!}{
\begin{tabular}{lrrr}
\toprule
 & \multicolumn{3}{c}{Corresponding refined mesh} \\  
 &  $256^3$  & $512^3 $ & $1,024^3$   \\
 \cmidrule{2-4}
\# cells  
&  $ 5,917,288$ & $ 27,444,592$ &  $122,076,544$ \\
\% cells  
&  $ 35.3$ & $ 20.4$ &  $11.4$ \\
\# blocks  
&  $148$ & $894$ &  $2,502$ \\
CPU time (s) 
&  $1,468$ & $ 11,350$ &  $117,731$ \\
\bottomrule
\end{tabular}
}
\end{center}
\end{table}

\section{Conclusions}
\label{secConclusions}
We presented and benchmarked a parallel solver with dynamic mesh adaptation for magnetohydrodynamics, implemented into the MPI-parallel distributed memory framework AMROC. The GLM-MHD model is discretised using finite volumes in two- and three-dimensional Cartesian geometries with explicit time integration. 
MR criteria are employed for triggering the mesh refinement.
We considered two classical MHD benchmarks, the Orszag--Tong vortex and the
magnetic shock cloud configuration. 
The accuracy and CPU time of the developed code were assessed and parallelisation issues including load balancing were analysed.
We showed that in comparison with the scaled gradient criterion the new MR implementation yields much better results in terms of accuracy and memory compression.
Moreover, our implementation presented a significant improvement compared to its MR-MHD base serial code.

This improvement is in agreement with the conclusions drawn in \cite{Deiterdingetal2016SIAM} for hydrodynamic problems where the total CPU time was the primary concern and it was found that \textcolor{black}{patch}-based hierarchical data structures yield a better choice. 
These data structures preserve some memory coherence on the computing data and using auxiliary data avoids repeated generation of topological and numerical procedures. 
However, this advantage comes at the cost of a more complex implementation. 
Yet, by using AMROC, including the verified \textcolor{black}{patch}-based AMR algorithm implementation plus parallelisation \cite{DeiterdingESAIM:2011}, and incorporating the MHD-GLM method as a patch integrator and implementing MR mesh adaptation as a refinement criterion, a performance enhanced two- and three-dimensional adaptive parallel MHD solver based on multiresolution principles has been realised in minimal time. 

\section*{Acknowledgements}

The authors thank the FAPESP SPRINT -- University of Southampton (Grant:16/ 50016-9), FAPESP (Grant: 2015/ 25624-2), CNPq (Grants: 307083/2017-9, 306038/2015-3, 140626/ 2014-0, 141741/2013-9), and FINEP (Grant: 0112052700) for financial support of this research. 
Also ML is grateful to CAPES and University of Southampton (SOU) for  financial support for his PhD internships and  SOU  and  I2M, Aix-Marseille Universit\'e for kind hospitality.
KS acknowledges financial support from the ANR-DFG, grant AIFIT (Grant 15-CE40-0019), and support by the French Research Federation for Fusion Studies (FRCM) within the framework of the European Fusion Development Agreement (EFDA). 
We are indebted to Eng. V. E. Menconi for his valuable computational assistance.

\section*{References}
\bibliographystyle{elsarticle-num} 
\bibliography{references}

\begin{thebibliography}{10}
\expandafter\ifx\csname url\endcsname\relax
  \def\url#1{\texttt{#1}}\fi
\expandafter\ifx\csname urlprefix\endcsname\relax\def\urlprefix{URL }\fi
\expandafter\ifx\csname href\endcsname\relax
  \def\href#1#2{#2} \def\path#1{#1}\fi

\bibitem{Schrijver20152745}
C.~J. Schrijver, K.~Kauristie, A.~D. Aylward, C.~M. Denardini, S.~E. Gibson,
  A.~Glover, N.~Gopalswamy, M.~Grande, M.~Hapgood, D.~Heynderickx, N.~Jakowski,
  V.~V. Kalegaev, G.~Lapenta, J.~A. Linker, S.~Liu, C.~H. Mandrini, I.~R. Mann,
  T.~Nagatsuma, D.~Nandy, T.~Obara, T.~P. O’Brien, T.~Onsager, H.~J.
  Opgenoorth, M.~Terkildsen, C.~E. Valladares, N.~Vilmer, Understanding space
  weather to shield society: A global road map for 2015–2025 commissioned by
  {COSPAR} and {ILWS}, Adv. Space Res. 55~(12) (2015) 2745 -- 2807.

\bibitem{Toth2012870}
G.~T\'oth, B.~van~der Holst, I.~V. Sokolov, D.~L. De~Zeeuw, T.~I. Gombosi,
  F.~Fang, Adaptive numerical algorithms in space weather modeling, J. Comput.
  Phys. 231~(3) (2012) 870--903.

\bibitem{Bittencourt:2004}
J.~Bittencourt, Fundamentals of Plasma Physics, Springer, 2004.

\bibitem{Gomes2015199}
A.~K.~F. Gomes, M.~O. Domingues, K.~Schneider, O.~Mendes, R.~Deiterding, An
  adaptive multiresolution method for ideal magnetohydrodynamics using
  divergence cleaning with parabolic-hyperbolic correction, Appl. Numer. Math.
  95 (2015) 199--213.

\bibitem{DominguesetalESAIM:2013}
M.~O. Domingues, A.~K.~F. Gomes, S.~M. Gomes, O.~Mendes, B.~Di~Pierro,
  K.~Schneider, Extended generalized {Lagrangian} multipliers for
  magnetohydrodynamics using adaptive multiresolution methods, ESAIM: Proc. 43
  (2013) 95--107.

\bibitem{Hejazialhosseini20108364}
B.~Hejazialhosseini, D.~Rossinelli, M.~Bergdorf, P.~Koumoutsakos, High order
  finite volume methods on wavelet-adapted grids with local time-stepping on
  multicore architectures for the simulation of shock-bubble interactions,
  Journal of Computational Physics 229~(22) (2010) 8364 -- 8383.

\bibitem{DeiterdingDominguesGomesRousselSchneiderESIAM:2009}
R.~Deiterding, M.~O. Domingues, S.~M. Gomes, O.~Roussel, K.~Schneider, Adaptive
  multiresolution or adaptive mesh refinement? a case study for {2D Euler}
  equations, {ESAIM} Proceedings 16 (2009) 181--194.

\bibitem{Deiterdingetal2016SIAM}
R.~Deiterding, M.~O. Domingues, S.~M. Gomes, K.~Schneider, Comparison of
  adaptive multiresolution and adaptive mesh refinement applied to simulations
  of the compressible {Euler} equations, {SIAM} Journal on Scientific Computing
  38~(5) (2016) S173--S193.

\bibitem{Deiterding-PhDThesis}
R.~Deiterding, Parallel adaptive simulation of multi-dimensional detonation
  structures, Ph.D. thesis, Brandenburgische Technische Universit\"at Cottbus
  (Sep 2003).

\bibitem{DeiterdingESAIM:2011}
R.~Deiterding, Block-structured adaptive mesh refinement - theory,
  implementation and application, ESAIM: Proc. 34 (2011) 97--150.

\bibitem{MoreiraLopesetal2018CAF}
M.~Moreira~Lopes, R.~Deiterding, A.~K.~F. Gomes, O.~Mendes, M.~O. Domingues, An
  ideal compressible magnetohydrodynamic solver with parallel block-structured
  adaptive mesh refinement, Comput. Fluids 173 (2018) 293--298.

\bibitem{Kallenrode:2004}
M.-B. Kallenrode, Space Physics An Introduction to Plasmas and Particles in the
  Heliosphere and Magnetospheres, Germany, Springer-Verlag, 2004.

\bibitem{KivelsonRussell:1996}
M.~G. Kivelson, C.~T. Russell (Eds.), Introduction to Space Physics, Cambridge,
  Cambridge University, 1996.

\bibitem{Cravens:2004}
T.~E. Cravens, Physics of Solar System Plasmas, Cambridge, Cambridge
  University, 2004.

\bibitem{RussellEtAl:2016}
C.~T. Russell, J.~G. Luhmann, R.~J. Strangeway, Space Physics An Introduction,
  Cambridge, Cambridge University, 2016.

\bibitem{BaumjohannTreumann:1999}
W.~Baumjohann, R.~A. Treumann, Basic Space Plasma Physics, London, Imperial
  College, 1999.

\bibitem{OrszagTang:1979}
S.~A. Orszag, C.-M. Tang, Small-scale structure of two-dimensional
  magnetohydrodynamic turbulence, J. Fluid Mech. 90~(1) (1979) 129--143.

\bibitem{BaloghTreumann:2013}
A.~Balogh, , R.~A. Treumann, Physics of Collisionless Shocks Space Plasma Shock
  Waves, New York, Springer, 2013.

\bibitem{Hopkins:2016}
P.~F. Hopkins, A constrained-gradient method to control divergence errors in
  numerical {MHD}, Mon. Notices Royal Astron. Soc. 462~(11) (2016) 576--587.

\bibitem{Dedneretal:2002}
A.~Dedner, F.~Kemm, D.~Kr{\"o}ner, C.-D. Munz, T.~Schnitzer, M.~Wesenberg,
  Hyperbolic divergence cleaning for the {MHD} equations, J. Comput. Phys.
  175~(2) (2002) 645--673.

\bibitem{mignone2010second}
A.~Mignone, P.~Tzeferacos, A second-order unsplit {G}odunov scheme for
  cell-centered {MHD}: The {CTU-GLM} scheme, J. Comput. Phys. 229~(6) (2010)
  2117--2138.

\bibitem{Leveque:2002}
R.~J. Leveque, Finite Volume Methods for Hyperbolic Systems, Cambridge
  University, 2002.

\bibitem{Berger-Oliger-84}
M.~J. Berger, J.~Oliger, Adaptive mesh refinement for hyperbolic partial
  differential equations, J. Comput. Phys. 53 (1984) 484--512.

\bibitem{Berger-Collela-88}
M.~Berger, P.~Colella, Local adaptive mesh refinement for shock hydrodynamics,
  J. Comput. Phys. 82 (1988) 64--84.

\bibitem{Bell-Berger-Saltzman-94}
J.~Bell, M.~Berger, J.~Saltzmann, M.~Welcome, Three-dimensional adaptive mesh
  refinement for hyperbolic conservation laws, {SIAM} J. Sci. Comput. 15 (1994)
  127--138.

\bibitem{DominguesetalESAIM:2011}
M.~O. Domingues, S.~M. Gomes, O.~Roussel, K.~Schneider, Adaptive
  multiresolution methods, ESAIM: Proc. 34 (2011) 1--96.

\bibitem{mueller:2001}
S.~{M\"uller}, Adaptive Multiscale Schemes for Conservation Laws, Vol.~27 of
  Lecture Notes in Computational Science and Engineering, Springer-Verlag,
  Heidelberg, 2003.

\bibitem{Harten:1995}
A.~Harten, Multiresolution algorithms for the numerical solution of hyperbolic
  conservation laws, Comm. Pure Appl. Math. 48 (1995) 1305--1342.

\bibitem{Harten:1996}
A.~Harten, Multiresolution representation of data: a general framework, {SIAM}
  {J. Numer. Anal.} 33~(3) (1996) 385--394.

\bibitem{CohenKaberMUllerPostel:2003}
A.~Cohen, S.~M. Kaber, S.~M\"uller, M.~Postel, Fully adaptive multirresolution
  finite volume schemes for conservation laws, Math. Comp. 72 (2003) 183--225.

\bibitem{Cohen:2000}
A.~Cohen, Wavelet methods in numerical analysis, in: P.~G. Ciarlet, J.~L. Lions
  (Eds.), Handbook of {N}umerical {A}nalysis, Vol. {VII}, Elsevier, Amsterdam,
  2000.

\bibitem{DominguesGomesRousselSchneider:APNUM2009}
M.~O. Domingues, S.~M. Gomes, O.~Roussel, K.~Schneider, Space-time adaptive
  multiresolution methods for hyperbolic conservation laws: Applications to
  compressible {Euler} equations, Appl. Numer. Math. 59~(9) (2009) 2303--2321.

\bibitem{Belletall:1994}
J.~Bell, M.~Berger, J.~Saltzmann, M.~Welcome, Three-dimensional adaptive mesh
  refinement for hyperbolic conservation laws, {SIAM} J. Sci. Comput. 15 (1994)
  127.

\bibitem{Deiterding-05}
R.~Deiterding, Construction and application of an {AMR} algorithm for
  distributed memory computers, in: T.~Plewa, T.~Linde, V.~G. Weirs (Eds.),
  Adaptive Mesh Refinement - Theory and Applications, Springer, 2005, pp.
  361--372.

\bibitem{MiyoshiKusano:2005}
T.~Miyoshi, K.~A. Kusano, A multi-state {HLL} approximate {Riemann} solver for
  ideal magnetohydrodynamics, J. Comput. Phys. 208 (2005) 315--344.

\bibitem{Toro:1999}
E.~F. Toro, Riemann Solvers and Numerical Methods for Fluid Dynamics, Springer,
  1999.

\bibitem{LandauLifshitz:2004}
L.~D. Landau, E.~M. Lifshitz, L.~P. Pitaevskii, Electrodynamics of Continuous
  Media, 2nd Edition, Vol. Volume 8 of Course of Theoretical Physics S,
  Pergamon, 2004.

\bibitem{Zacharyetal:1994}
A.~Zachary, A.~Malagoli, P.~Colella, A higher-order {Godunov} method for
  multidimensional ideal magnetohydrodynamics, {SIAM} Journal on Scientific
  Computing 15~(2) (1994) 263--284.

\bibitem{londrillo2000high}
P.~Londrillo, L.~Del~Zanna, High-order upwind schemes for multidimensional
  magnetohydrodynamics, Astrophys. J. 530~(1) (2000) 508--524.

\bibitem{DEITERDING:2019CAFMRSG}
R.~Deiterding, M.~O. Domingues, K.~Schneider, Multiresolution analysis as a
  criterion for effective dynamic mesh adaptation - a case study for {Euler
  equations in the SAMR framework AMROC}., Computer \& Fluids(submitted).

\bibitem{TOUMA2006617}
R.~Touma, P.~Arminjon, Central finite volume schemes with constrained transport
  divergence treatment for three-dimensional ideal {MHD}, Journal of
  Computational Physics 212~(2) (2006) 617 -- 636.

\bibitem{Gomes:2018:SiNuMo}
A.~K.~F. Gomes, Simula{\c{c}}{\~a}o num{\'e}rica de um modelo
  magneto-hidrodin{\^a}mico multidimensional no contexto da
  multirresolu{\c{c}}{\~a}o adaptativa por m{\'e}dias celulares, Ph.D. thesis,
  Instituto Nacional de Pesquisas Espaciais (INPE), S{\~a}o Jos{\'e} dos
  Campos, in Portuguese (Dec 2018).

\bibitem{RSTB03}
O.~Roussel, K.~Schneider, A.~Tsigulin, H.~Bockhorn, A conservative fully
  adaptive multiresolution algorithm for parabolic {PDEs}, J. Comput. Phys.
  188~(2) (2003) 493--523.

\end{thebibliography}

\end{document}